\documentclass[aps,twocolumn,prd,superscriptaddress]{revtex4}
\usepackage[dvips]{graphicx}
\usepackage{amsmath}
\usepackage{amssymb}
\newcommand{\ud}{\mathrm{d}}
\newcommand{\dirac}{\partial\llap{$\diagup$\kern-2pt}}
\newcommand{\fett}[1]{\boldsymbol{#1}}
\newcommand{\fettu}[1]{\mathbf{#1}}

\newcommand{\diag}{\mathrm{diag}}

\newcommand{\Tr}{\mathrm{Tr}}

\newcommand{\e}{\mathrm{e}}
\begin{document}

\title{The phase diagram of neutral quark matter: \\
Self-consistent treatment of quark masses}

%%%%%%%%%%%%% begin addresses %%%%%%%%%%%%%%%%%%%

\author{Stefan B.\ R\"uster}
\email{ruester@th.physik.uni-frankfurt.de}
\affiliation{Institut f\"ur Theoretische Physik, 
J.W.\ Goethe-Universit\"at,
D-60438 Frankfurt am Main, Germany}

\author{Verena Werth}
\email{verena.werth@physik.tu-darmstadt.de}
\affiliation{Institut f\"ur Kernphysik, 
Technische Universit\"at Darmstadt,
D-64289 Darmstadt, Germany}

\author{Michael Buballa}
\email{michael.buballa@physik.tu-darmstadt.de}
\affiliation{Institut f\"ur Kernphysik, 
Technische Universit\"at Darmstadt,
D-64289 Darmstadt, Germany}

\author{Igor A.\ Shovkovy}
\email{shovkovy@th.physik.uni-frankfurt.de}
  \altaffiliation[on leave from ]{%
       Bogolyubov Institute for Theoretical Physics,
       03143, Kiev, Ukraine}
\affiliation{
Frankfurt Institute for Advanced Studies, J.W.\ Goethe-Universit\"at,
D-60438 Frankfurt am Main, Germany}

\author{Dirk H.\ Rischke}
\email{drischke@th.physik.uni-frankfurt.de}
\affiliation{Institut f\"ur Theoretische Physik, 
J.W.\ Goethe-Universit\"at,
D-60438 Frankfurt am Main, Germany}

%%%%%%%%%%%%% end addresses %%%%%%%%%%%%%%%%%%%%%

\date{\today}

%%%%%%%%%%%%% begin abstract %%%%%%%%%%%%%%%%%%%%
\begin{abstract}
We study the phase diagram of dense, locally neutral three-flavor 
quark matter within the framework of the Nambu--Jona-Lasinio model. 
In the analysis, dynamically generated quark masses are taken into 
account self-consistently. The phase diagram in the plane of 
temperature and quark chemical potential is presented. The results
for two qualitatively different regimes, intermediate and strong 
diquark coupling strength, are presented. It is shown that the 
role of gapless phases diminishes with increasing diquark 
coupling strength.
\end{abstract}
%%%%%%%%%%%%% end abstract %%%%%%%%%%%%%%%%%%%%%%

\maketitle

%%%%%%%%%%%%%%%%%%%%%%%%%%%%%%%%%%%%%%%%%%%%%%%%%
%%%%%%%%%%%%%%%%%%%%%%%%%%%%%%%%%%%%%%%%%%%%%%%%% 
\section{introduction}

Theoretical studies suggest that baryon matter at sufficiently high 
density and sufficiently low temperature is a color superconductor. 
(For reviews on color superconductivity see, for example, 
Ref.~\cite{reviews}.) In Nature, the highest densities of matter are 
reached in central regions of compact stars. There, the density might
be as large as $10 \rho_{0}$ where $\rho_{0}\approx 0.15$ fm$^{-3}$ 
is the saturation density. It is possible that baryonic matter is 
deconfined under such conditions and, perhaps, it is 
color-superconducting. 

In compact stars, matter in the bulk is neutral with respect to the
electric and color charges. Matter should also remain in $\beta$ 
equilibrium. Taking these constraints consistently into account may 
have a strong effect on the competition between different phases 
of deconfined quark matter at large baryon densities 
\cite{absence2sc,SRP,mei,g2SC,gCFL}. In this paper, we study this 
competition within the framework of a Nambu--Jona-Lasinio (NJL) model. 
The results are summarized in the phase diagram in the plane of 
temperature ($T$) and quark chemical potential ($\mu$).

The first attempt to obtain the phase diagram of dense, 
locally neutral three-flavor quark matter as a function of the 
strange quark mass, the quark chemical potential, and the 
temperature was made in Ref.~\cite{phase-d}. It was shown that, 
at zero temperature and small values of the strange quark mass, 
the ground state of matter corresponds
to the color-flavor-locked (CFL) phase \cite{cfl,weakCFL}. At some 
critical value of the strange quark mass, this is replaced by the 
gapless CFL (gCFL) phase \cite{gCFL}. In addition, several other 
phases were found at nonzero temperature. For instance, it was 
shown that there should exist a metallic CFL (mCFL) phase, a 
so-called uSC phase \cite{dSC}, as well as the standard two-flavor 
color-superconducting (2SC) phase \cite{cs,weak} and the gapless 
2SC (g2SC) phase \cite{g2SC}. 

In Ref.~\cite{phase-d}, the effect of the strange quark mass 
was incorporated only approximately through a shift of the chemical
potential of strange quarks, $\mu_{s} \to \mu_s - m_s^2/(2\mu)$. 
While such an approach is certainly reliable at small values of the 
strange quark mass, it becomes uncontrollable with increasing the 
mass. The phase diagram of Ref.~\cite{phase-d} was further developed 
in Refs.~\cite{phase-d1,phase-d2} where the shift-approximation in 
dealing with the strange quark was not employed any more. So far, 
however, quark masses were treated as free parameters, rather 
than dynamically generated quantities. In this paper, we study 
the phase diagram of dense, locally neutral three-flavor quark
matter within the NJL model, treating dynamically generated quark 
masses self-consistently. Some results within this approach at zero 
temperature were also obtained in Refs.~\cite{SRP,kyoto}.

As in Refs.~\cite{phase-d,phase-d1,phase-d2}, we restrict our 
analysis to locally neutral phases only. This automatically 
excludes, for example, mixed \cite{mix} and crystalline \cite{cryst} 
phases. Taking them into account requires a special treatment which
is outside the scope of this paper. 

This paper is organized as follows. In Sec.~\ref{eos}, we introduce 
the model and, within this model, derive a complete set of gap 
equations and charge neutrality conditions.  
The numerical results for the phase diagrams in the plane of temperature 
and quark chemical potential in two qualitatively different regimes 
are presented in Sec.~\ref{results}. Finally, our results are 
summarized in Sec.~\ref{conclusions}. The Appendix contains some 
useful formulas.

%%%%%%%%%%%%%%%%%%%%%%%%%%%%%%%%%%%%%%%%%%%%
%%%%%%%%%%%%%%%%%%%%%%%%%%%%%%%%%%%%%%%%%%%%
\section{Model and formalism}
\label{eos}

In this paper, we use a three-flavor quark model with a local
NJL-type interaction. The Lagrangian density is given by 
\begin{eqnarray}
\mathcal{L} &=& \bar \psi \, ( i \dirac - \hat{m} \, ) \psi 
+G_S \sum_{a=0}^8 \left[ \left( \bar \psi \lambda_a \psi \right)^2 
+ \left( \bar \psi i \gamma_5 \lambda_a \psi \right)^2 \right] 
\nonumber \\
&+& G_D \sum_{\gamma,c} \left[\bar{\psi}_{\alpha}^{a} i \gamma_5
\epsilon^{\alpha \beta \gamma}
\epsilon_{abc} (\psi_C)_{\beta}^{b} \right] \left[ 
(\bar{\psi}_C)_{\rho}^{r} i \gamma_5
\epsilon^{\rho \sigma \gamma} \epsilon_{rsc} \psi_{\sigma}^{s} 
\right] 
\nonumber \\
&-& K \left\{ \det_{f}\left[ \bar \psi \left( 1 + \gamma_5 \right) \psi
\right] + \det_{f}\left[ \bar \psi \left( 1 - \gamma_5 \right) \psi
\right] \right\} \;,
\label{Lagrangian}
\end{eqnarray}
where the quark spinor field $\psi_{\alpha}^{a}$ carries color 
($a=r,g,b$) and flavor ($\alpha=u,d,s$) indices. The matrix of quark 
current masses is given by $\hat{m} = \diag_{f}(m_u, m_d, m_s)$.
Regarding other notations, $\lambda_a$ with $a=1,\ldots,8$ are 
the Gell-Mann matrices in flavor space, and $\lambda_0\equiv 
\sqrt{2/3} \,\openone_{f}$. The charge conjugate spinors are 
defined as follows: $\psi_C = C \bar \psi^T$ and $\bar 
\psi_C = \psi^T C$, where $\bar\psi=\psi^\dagger \gamma^0$ 
is the Dirac conjugate spinor and $C=i\gamma^2 \gamma^0$ 
is the charge conjugation matrix.

The model in Eq.~(\ref{Lagrangian}) should be viewed as an 
effective model of strongly interacting matter that captures at 
least some key features of QCD dynamics. The Lagrangian density 
contains three different interaction terms which are chosen to 
respect the symmetries of QCD. Note that we include the 't Hooft 
interaction whose strength is determined by the coupling constant 
$K$. This term breaks $U(1)$ axial symmetry.

The term in the second line of Eq.~(\ref{Lagrangian}) describes a scalar
diquark interaction in the color antitriplet and flavor antitriplet
channel. For symmetry reasons there should also be a pseudoscalar 
diquark interaction with the same coupling constant. This term 
would be important to describe Goldstone boson condensation in the 
CFL phase~\cite{goldstones}. In this paper, however, we neglect 
this possibility and therefore drop the pseudoscalar diquark term.

We use the following set of model parameters \cite{RKH}:
\begin{subequations}
\label{model-parameters}
\begin{eqnarray}
m_{u,d} &=& 5.5 \; \mathrm{MeV} \; , \\
m_s &=& 140.7 \; \mathrm{MeV} \; , \\
G_S \Lambda^2 &=& 1.835 \; , \\
K \Lambda^5 &=& 12.36 \; , \\
\Lambda &=& 602.3 \; \mathrm{MeV} \; .
\label{Lambda} 
\end{eqnarray}
\end{subequations}
After fixing the masses of the up and down quarks at equal values, 
$m_{u,d}=5.5~\mbox{MeV}$, the other four parameters are chosen to 
reproduce the following four observables of vacuum QCD \cite{RKH}: 
$m_{\pi}=135.0~\mbox{MeV}$, $m_{K}=497.7~\mbox{MeV}$,
$m_{\eta^\prime}=957.8~\mbox{MeV}$, and $f_{\pi}=92.4~\mbox{MeV}$.
This parameter set gives $m_{\eta}=514.8~\mbox{MeV}$ \cite{RKH}.

In Ref.~\cite{RKH}, the diquark coupling $G_D$ was not fixed by the 
fit of the meson spectrum in vacuum. In general, it is expected to be 
of the same order as the quark-antiquark coupling $G_S$. In this paper, 
we study in detail two possible cases: the regime of intermediate 
coupling strength with $G_D=\frac34 G_S$, and the regime of strong 
coupling with $G_D=G_S$. The comparison of phase diagrams in these 
two cases will turn out to be very instructive. 

The grand partition function, up to an irrelevant normalization 
constant, is given by 
\begin{equation}
\label{Z}
\mathcal{Z} \equiv \e^{-\Omega V/T}
= \int \mathcal{D} \bar\psi \mathcal{D} \psi \, \e^{i
\int_X \left( \mathcal{L} + \bar\psi \hat{\mu} \gamma^0 \psi
\right) } \; ,
\end{equation}
where $\Omega$ is the thermodynamic potential density, $V$ is the 
volume of the three-space, and $\hat\mu$ is a diagonal matrix of 
quark chemical potentials. In chemical equilibrium (which provides 
$\beta$ equilibrium as a special case), the nontrivial components 
of this matrix are extracted from the following relation:
\begin{equation}
\mu_{ab}^{\alpha\beta} = \left(
  \mu \delta^{\alpha\beta} 
+ \mu_Q Q_{f}^{\alpha\beta} \right)\delta_{ab} 
+ \left[ \mu_3 \left(T_3\right)_{ab} 
+ \mu_8 \left(T_8\right)_{ab} \right] \delta^{\alpha\beta} \; .
\label{mu-f-i}
\end{equation}
Here $\mu$ is the quark chemical potential (by definition, 
$\mu = \mu_{B}/3$ where $\mu_{B}$ is the baryon chemical 
potential), $\mu_Q$ is the chemical potential of electric charge, 
while $\mu_3$ and $\mu_8$ are color chemical potentials associated 
with two mutually commuting color charges of the $SU(3)_c$ gauge 
group. The explicit form of the electric charge matrix is 
        $Q_{f}=\mbox{diag}_{f}(\frac23,-\frac13,-\frac13)$,
and the explicit form of the color charge matrices is
        $T_3=\mbox{diag}_c(\frac12,-\frac12,0)$ and 
$\sqrt{3}T_8=\mbox{diag}_c(\frac12,\frac12,-1)$.

In order to calculate the mean-field thermodynamic potential 
at temperature $T$, we first linearize the interaction in the 
presence of the diquark condensates $\Delta_{c} \sim 
(\bar{\psi}_C)_{\alpha}^{a} i \gamma_5 \epsilon^{\alpha \beta c} 
\epsilon_{a b c} \psi_{\beta}^{b}$ (no sum over $c$) and the 
quark-antiquark condensates $\sigma_\alpha \sim \bar \psi_\alpha^a 
\psi_\alpha^a$ (no sum over $\alpha$). Then, integrating out the 
quark fields and neglecting the fluctuations of composite order 
parameters, we arrive at the following expression for the 
thermodynamic potential:
\begin{eqnarray}
\Omega &=& \Omega_{L} +
\frac{1}{4 G_D} \sum_{c=1}^{3} \left| \Delta_c \right|^2
+2 G_S \sum_{\alpha=1}^{3} \sigma_\alpha^2 \nonumber\\
&-& 4 K \sigma_u \sigma_d \sigma_s
-\frac{T}{2V} \sum_K \ln \det \frac{S^{-1}}{T} \; ,
\label{Omega}
\end{eqnarray}
where we also added the contribution of leptons, $\Omega_{L}$, 
which will be specified later. 

We should note that we have restricted ourselves to field contractions
corresponding to the Hartree approximation. 
In a more complete treatment, among others, 
the 't Hooft interaction term gives also rise 
to mixed contributions containing both diquark and quark-antiquark 
condensates, i.e., $\propto \sum_{\alpha=1}^{3} \sigma_\alpha 
|\Delta_\alpha|^2$~\cite{RSSV00}. In this study, as in Refs.~\cite{SRP,MB}, 
we neglect such terms for simplicity. While their presence may change 
the results quantitatively, one does not expect them to modify the 
qualitative structure of the phase diagram. 

In Eq.~(\ref{Omega}), $S^{-1}$ is the inverse full quark propagator 
in the Nambu-Gorkov representation, 
\begin{equation}
S^{-1} = 
\left(
\begin{array}{cc}
[ G_0^+ ]^{-1} & \Phi^- \\
\Phi^+ & [ G_0^- ]^{-1}
\end{array}
\right) \; ,
\label{off-d}
\end{equation}
with the diagonal elements being the inverse Dirac 
propagators of quarks and of charge-conjugate quarks, 
\begin{equation}
[ G_0^\pm ]^{-1} = \gamma^\mu K_\mu 
\pm \hat \mu \gamma_0 - \hat{M} \; ,
\end{equation}
where $K^\mu = (k_0, \fettu{k})$ denotes the four-momentum of the 
quark. At nonzero temperature, we use the Matsubara imaginary time 
formalism. Therefore, the energy $k_0$ is replaced with $-i \omega_n$ 
where $\omega_n\equiv (2n+1)\pi T$ are the fermionic Matsubara 
frequencies. Accordingly, the sum over $K$ in Eq.~(\ref{Omega}) 
should be interpreted as a sum over integer $n$ and an integral 
over the three-momentum $\fettu{k}$.

The constituent quark mass matrix is defined as 
$\hat{M} = \diag_{f}(M_u,M_d,M_s)$ with 
\begin{equation}
M_\alpha = m_\alpha - 4 G_S \sigma_\alpha 
+ 2 K \sigma_\beta \sigma_\gamma \; ,
\label{Mi}
\end{equation}
where $\sigma_\alpha $ are the quark-antiquark condensates, and
the set of indices $(\alpha, \beta, \gamma)$ is a permutation of 
$(u,d,s)$.

The off-diagonal components of the propagator (\ref{off-d}) 
are the so-called gap matrices given in terms of three diquark
condensates. The color-flavor structure of these matrices is 
given by
\begin{equation}
\left(\Phi^-\right)^{\alpha\beta}_{ab} 
= -\sum_c \epsilon^{\alpha\beta c}\,\epsilon_{abc}
   \,\Delta_c\,\gamma_5~,
\label{Phim}
\end{equation}
and $\Phi^+ = \gamma^0 (\Phi^-)^\dagger \gamma^0$. Here, as before, 
$a$ and $b$ refer to the color components and $\alpha$ and $\beta$ 
refer to the flavor components. Hence, the gap parameters $\Delta_1$, 
$\Delta_2$, and $\Delta_3$ correspond to the down-strange, the up-strange 
and the up-down diquark condensates, respectively. All three of them 
originate from the color-antitriplet, flavor-antitriplet diquark 
pairing channel. For simplicity, the color and flavor symmetric 
condensates are neglected in this study. They were shown to be 
small and not crucial for the qualitative understanding of the 
phase diagram \cite{phase-d}.

By making use of the results in Appendix~\ref{tpot}, the determinant
of the inverse quark propagator can be decomposed as follows:
\begin{equation}
   \det \frac{S^{-1}}{T}
=  \prod_{i=1}^{18} 
   \left( \frac{ \omega_n^2 + \epsilon_i ^2 }{T^2}
   \right)^2 \; ,
\label{det-S}
\end{equation}
where $\epsilon_i$ are eighteen independent positive energy eigenvalues. 
The Matsubara summation in Eq.~(\ref{Omega}) can then be done analytically 
by employing the relation \cite{Kapusta}
\begin{equation}
\sum_n \ln \left( \frac{ \omega_n^2 + \epsilon_i^2}{T^2} \right) 
= \frac{| \epsilon_i|}{T} 
+ 2 \ln \left( 1 + \e^{- \frac{| \epsilon_i |}{T}} \right) \; .
\end{equation}
Then, we arrive at the following mean-field expression for the 
pressure ($p\equiv -\Omega$):
\begin{eqnarray}
p &=& \frac{1}{2 \pi^2} \sum_{i=1}^{18} \int_0^\Lambda \ud k \, k^2
\left[ |\epsilon_i| + 2 T \ln \left( 1 + \e^{-
\frac{|\epsilon_i|}{T}} \right) \right] \nonumber \\
&+& 4 K \sigma_u \sigma_d \sigma_s
- \frac{1}{4 G_D} \sum_{c=1}^{3} \left| \Delta_c \right|^2
-2 G_S \sum_{\alpha=1}^{3} \sigma_\alpha^2
\nonumber \\
&+& \frac{T}{\pi^2} \sum_{l=e,\mu} \sum_{\epsilon=\pm}
\int_0^\infty \ud k \, k^2
\ln \left( 1 + \e^{-\frac{E_l-\epsilon\mu_l}{T}}\right)\; ,
\label{pressure}
\end{eqnarray}
where the contribution of electrons and muons with masses $m_e \approx 
0.511$ MeV and $m_\mu \approx 105.66$ MeV were included. Note that muons
may exist in matter in $\beta$ equilibrium and, therefore, they are 
included in the model for consistency. However, being about 200 times 
heavier than electrons, they do not play a big role in the analysis.

In this paper, we assume that there are no trapped neutrinos in 
quark matter. This is expected to be a good approximation for matter 
inside a neutron star after the short deleptonization period is over. 
The effect of neutrino trapping will be addressed elsewhere \cite{pd-nu}.

The expression for the pressure in Eq.~(\ref{pressure}) has a physical 
meaning only when the chiral and color superconducting order parameters,
$\sigma_\alpha$ and $\Delta_c$, satisfy the following set of six gap equations:
\begin{subequations}
\label{gapeqns}
\begin{eqnarray}
\frac{\partial p}{\partial \sigma_\alpha} &=& 0 \;  , \\
\frac{\partial p}{\partial \Delta_c} &=& 0 \; .
\end{eqnarray}
\end{subequations}
To enforce the conditions of local charge neutrality in dense matter,
we also require three other equations to be satisfied,
\begin{subequations}
\label{neutrality}
\begin{eqnarray}
n_Q &\equiv& \frac{ \partial p }{\partial \mu_Q} = 0 \; , \\
n_3 &\equiv& \frac{ \partial p }{\partial \mu_3} = 0 \; , \\
n_8 &\equiv& \frac{ \partial p }{\partial \mu_8} = 0 \; .
\end{eqnarray}
\end{subequations}
These fix the values of the three corresponding chemical potentials,
$\mu_Q$, $\mu_3$ and $\mu_8$. After these are fixed, only the quark 
chemical potential $\mu$ is left as a free parameter.

%%%%%%%%%%%%%%%%%%%%%%%%%%%%%%%%%%%%%%%%%%%%
%%%%%%%%%%%%%%%%%%%%%%%%%%%%%%%%%%%%%%%%%%%%
\section{Results}
\label{results}

In order to obtain the phase diagram, we have to find the ground 
state of matter for each given set of the parameters in the model.
In the case of locally neutral matter, there are two parameters
that should be specified: temperature $T$ and quark chemical potential 
$\mu$. After these are fixed, one has to compare the values of the 
pressure in all competing neutral phases of quark matter. The ground 
state corresponds to the phase with the highest pressure. 

Before calculating the pressure, given by Eq.~(\ref{pressure}), one 
has to find the values of the chiral and the color superconducting 
order parameters, $\sigma_\alpha$ and $\Delta_c$, as well as the values 
of the three charge chemical potentials, $\mu_Q$, $\mu_3$ and $\mu_8$.
These are obtained by solving the coupled set of six gap equations 
(\ref{gapeqns}) together with the three neutrality conditions 
(\ref{neutrality}). By using standard numerical recipes, it is 
not extremely difficult to find a solution to the given set of 
nine nonlinear equations. Complications arise, however, due to 
the fact that often the solution is not unique. 

The existence of different solutions to the same set of equations, 
(\ref{gapeqns}) and (\ref{neutrality}), reflects the physical 
fact that there could exist several competing neutral phases with 
different physical properties. Among these phases, all but one 
are unstable or metastable. In order to take this into account 
in our study, we look for the solutions of the following 8 types:

\begin{enumerate}

\item Normal quark (NQ) phase: $\Delta_1=\Delta_2=\Delta_3=0$;

\item 2SC phase: $\Delta_1=\Delta_2=0$,  and $\Delta_3\neq 0$;

\item 2SC${us}$ phase: $\Delta_1=\Delta_3=0$,  and $\Delta_2\neq 0$;

\item 2SC${ds}$ phase: $\Delta_2=\Delta_3=0$,  and $\Delta_1\neq 0$;

\item uSC phase:  $\Delta_2\neq 0$, $\Delta_3\neq 0$, and $\Delta_1= 0$;

\item dSC phase: $\Delta_1\neq 0$, $\Delta_3\neq 0$, and $\Delta_2= 0$;

\item sSC phase: $\Delta_1\neq 0$, $\Delta_2\neq 0$, and $\Delta_3= 0$;

\item CFL phase: $\Delta_1\neq0$, $\Delta_2\neq 0$, $\Delta_3\neq 0$.

\end{enumerate}

Then, we calculate the values of the pressure in all nonequivalent 
phases, and determine the ground state as the phase with the highest 
pressure. After this is done, we study additionally the spectrum of 
low-energy quasiparticles in search for the existence of gapless modes. 
This allows us to refine the specific nature of the ground state. 

In the above definition of the eight phases in terms of $\Delta_c$,
we have ignored the quark-antiquark condensates $\sigma_\alpha$.
In fact, in the chiral limit ($m_\alpha= 0$), the quantities 
$\sigma_\alpha$ are good order parameters and we could define 
additional sub-phases characterized by nonvanishing values of 
one or more $\sigma_\alpha$. With the model parameters at hand, 
however, chiral symmetry is broken explicitly by the nonzero current 
quark masses, and the values of $\sigma_\alpha$ never vanish. 
Hence, in a strict sense it is impossible to define any new phases 
in terms of $\sigma_\alpha$. 

Of course, this does not exclude the possibility of discontinuous 
changes in $\sigma_\alpha$ at some line in the plane of temperature 
and quark chemical potential, thereby constituting a first-order 
phase transition line. It is generally expected that the ``would-be'' 
chiral phase transition remains first-order at low temperatures, 
even for nonzero quark masses. Above some critical temperature, 
however, this line could end in a critical endpoint and there is 
only a smooth crossover at higher temperatures. Among others, this 
picture emerges from NJL-model studies, both, without~\cite{cr-pt-NJL} 
and with~\cite{BO} diquark pairing (see also Ref.~\cite{MB}). We 
should therefore expect a similar behavior in our analysis.

\begin{figure*}[h!]
%%%% h - here, t - top, b - bottom, p - page, ! - as soon as possible 
  \begin{center}
    \includegraphics[width=0.75\textwidth]{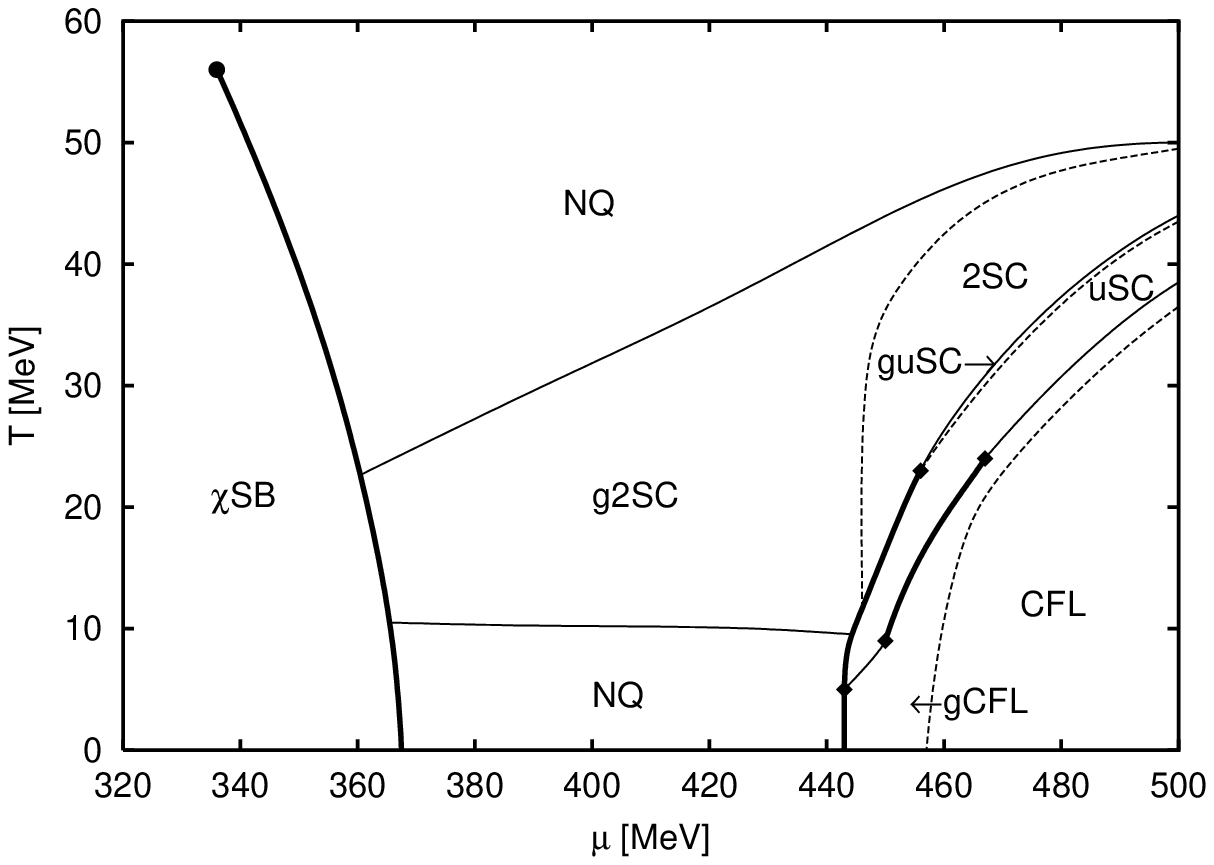}
    \caption{The phase diagram of neutral quark matter in the regime
     of intermediate diquark coupling strength, $G_{D}=\frac34 G_{S}$.
     First-order phase boundaries are indicated by bold solid lines,
     whereas the thin solid lines mark second-order phase boundaries
     between two phases which differ by one or more nonzero diquark
     condensates. The dashed lines indicate the (dis-)appearance of 
     gapless modes in different phases, and they do not correspond 
     to phase transitions.}
    \label{phasediagram}
  \end{center}
\end{figure*}

\begin{figure*}[ht]
%%%% h - here, t - top, b - bottom, p - page, ! - as soon as possible 
  \begin{center}
    \includegraphics[width=0.75\textwidth]{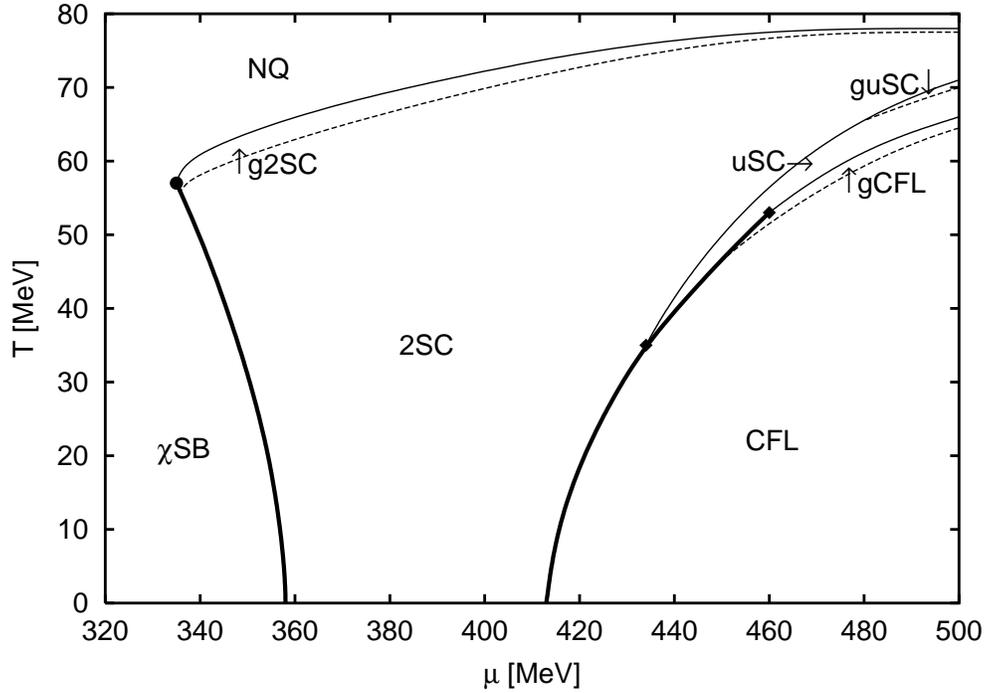}
    \caption{The phase diagram of neutral quark matter in the regime
     of strong diquark coupling, $G_{D}=G_{S}$. The meaning of the 
     various line types is the same as in Fig.~\ref{phasediagram}.}
    \label{phasediagram-strong}
  \end{center}
\end{figure*}

\begin{figure*}[h!]
%%%% h - here, t - top, b - bottom, p - page, ! - as soon as possible 
  \begin{center}
    \includegraphics[width=0.95\textwidth]{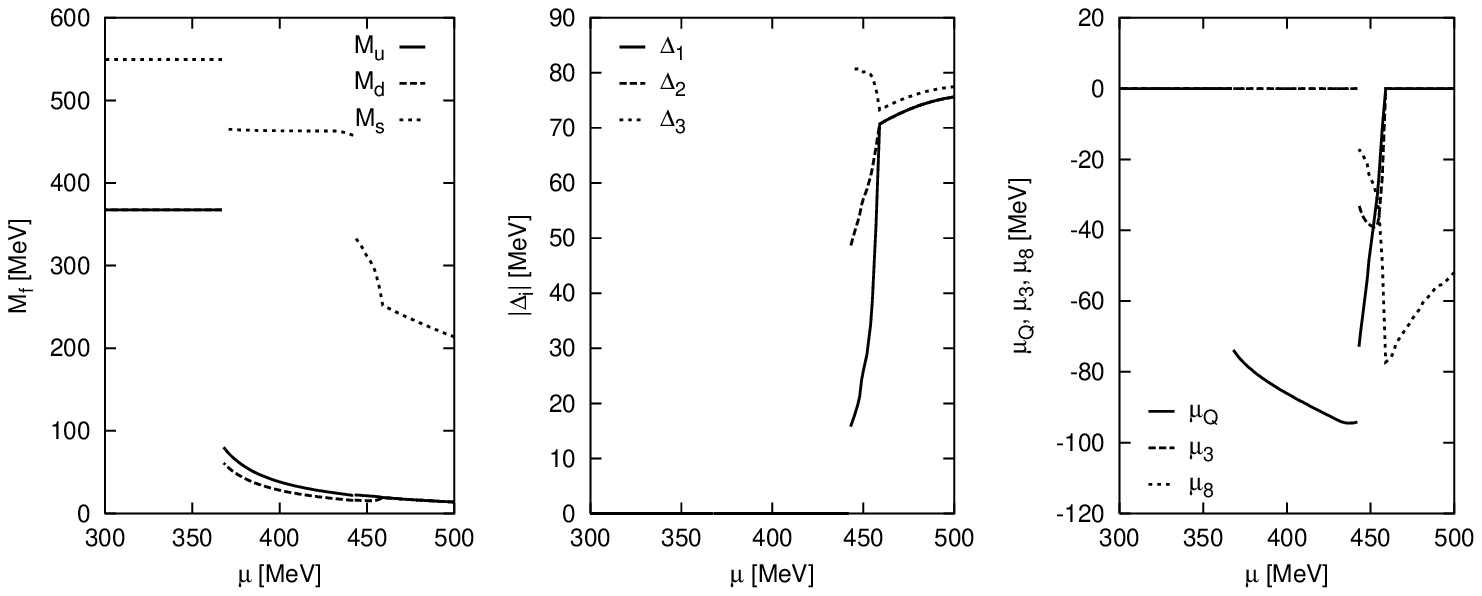}\\
    \includegraphics[width=0.95\textwidth]{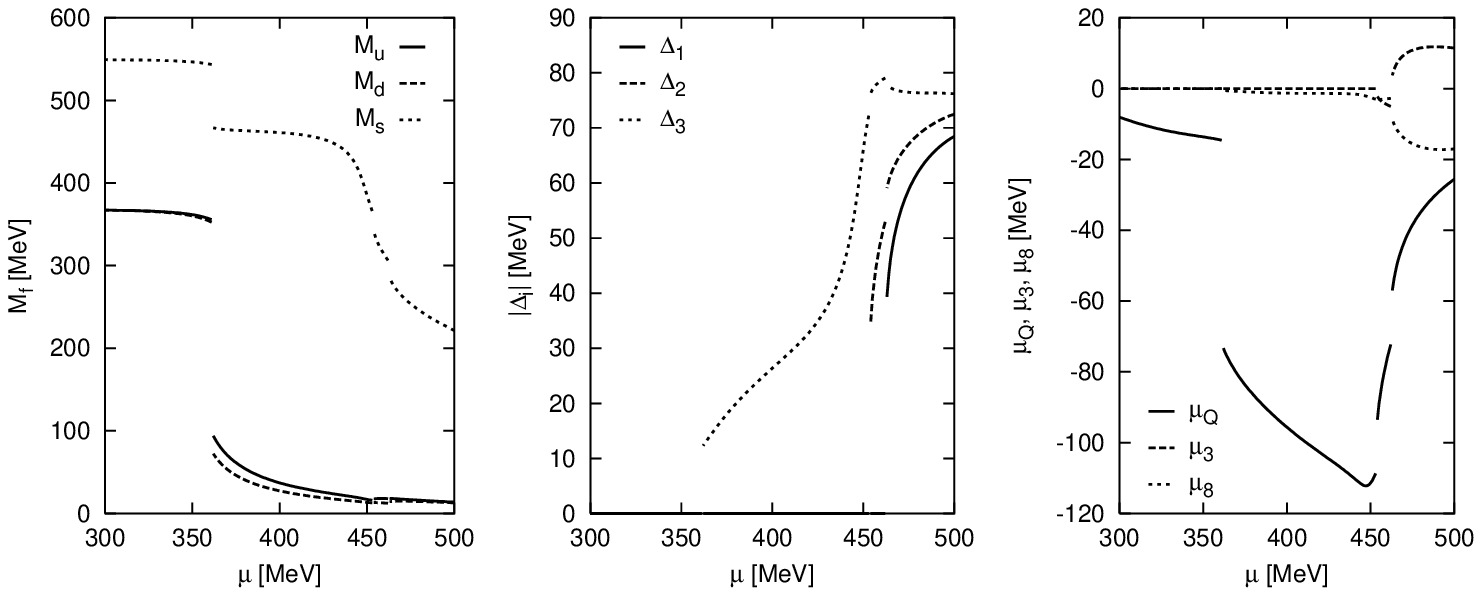}\\
    \includegraphics[width=0.95\textwidth]{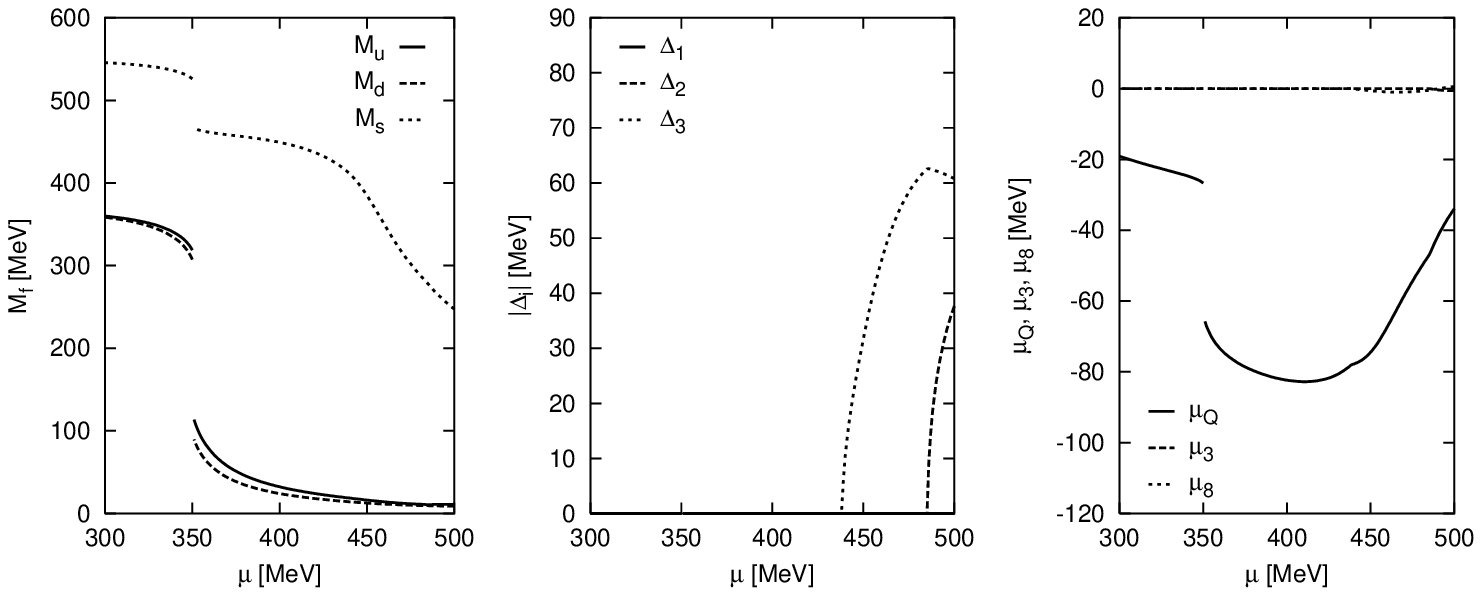}
    \caption{Dependence of the quark masses, of the gap parameters, 
     and of the electric and color charge chemical potentials on the 
     quark chemical 
     potential at a fixed temperature, $T=0~\mbox{MeV}$ (three upper 
     panels), $T=20~\mbox{MeV}$ (three middle panels), and $T=40~\mbox{MeV}$ 
     (three lower panels). The diquark coupling strength is $G_D=\frac34 G_S$.}
    \label{plot0-20-40}
  \end{center}
\end{figure*}

\begin{figure*}[h!]
%%%% h - here, t - top, b - bottom, p - page, ! - as soon as possible 
  \begin{center}
    \includegraphics[width=0.95\textwidth]{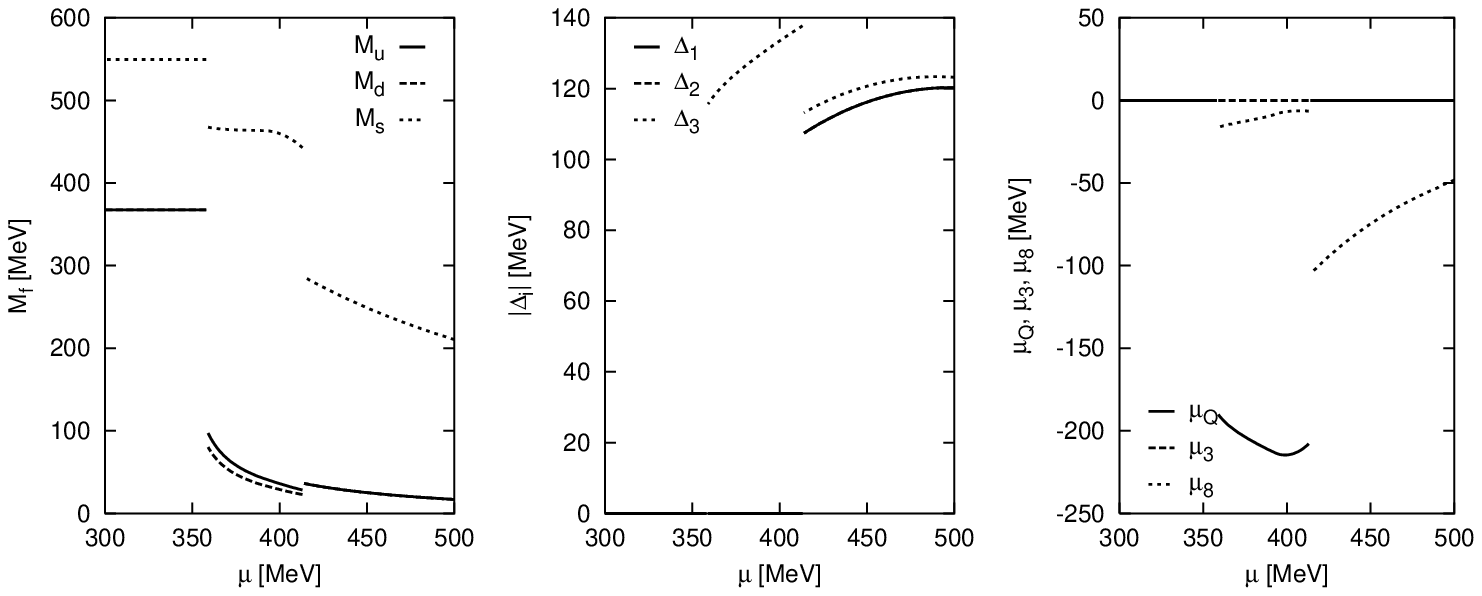}\\
    \includegraphics[width=0.95\textwidth]{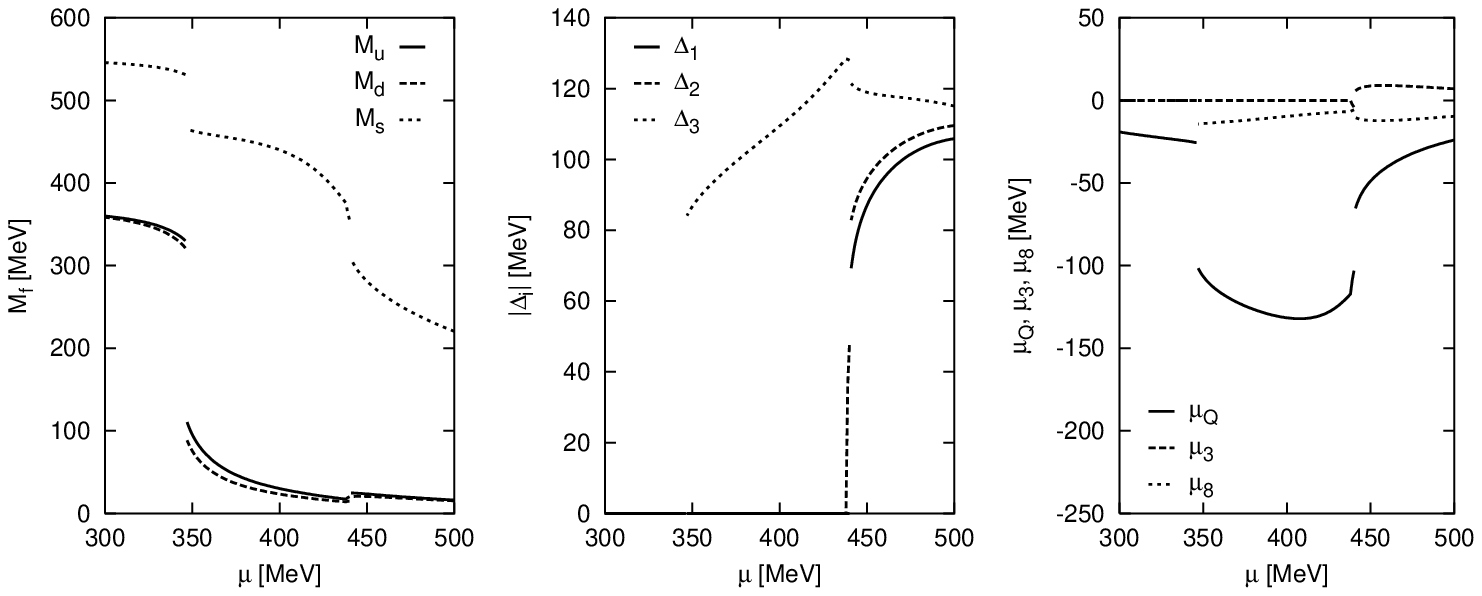}\\
    \includegraphics[width=0.95\textwidth]{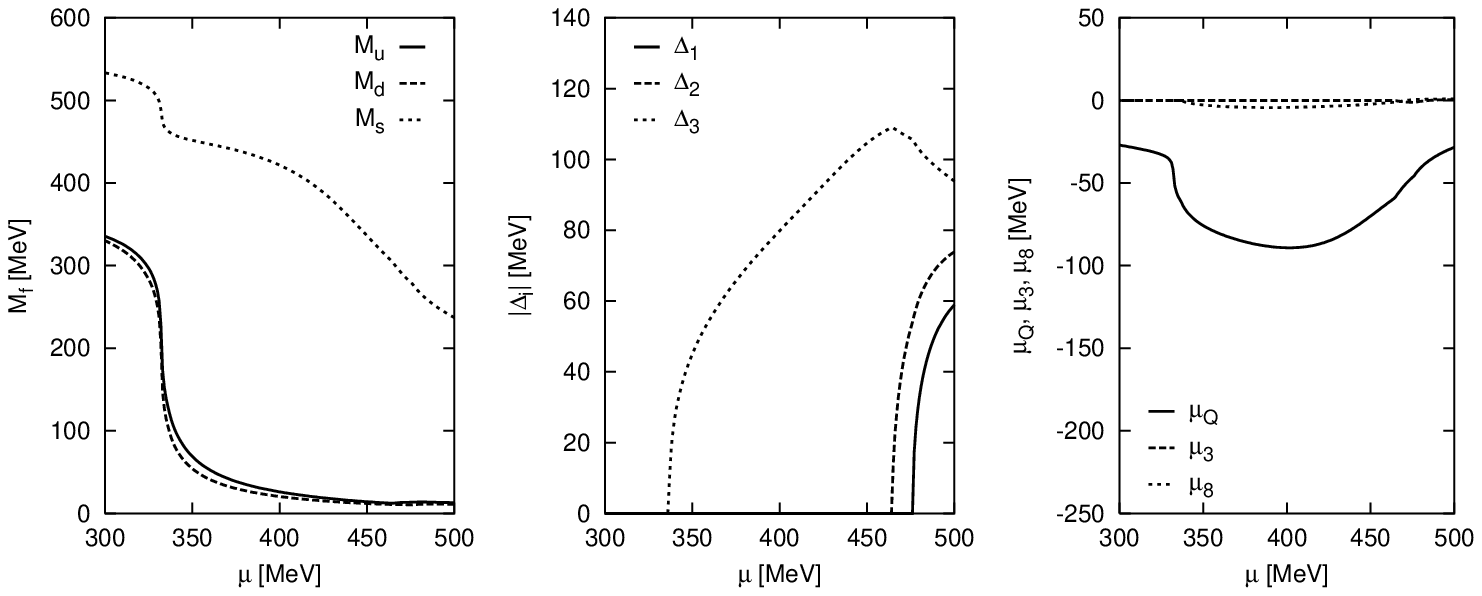}
    \caption{The dependence of the quark masses, of the gap parameters 
     and of the charge chemical potentials on the quark chemical 
     potential at a fixed temperature, $T=0~\mbox{MeV}$ (three upper 
     panels), $T=40~\mbox{MeV}$ (three middle panels), and $T=60~\mbox{MeV}$ 
     (three lower panels). The diquark coupling strength is $G_D=G_S$.}
    \label{plot0-40-60-strong}
  \end{center}
\end{figure*}

Our numerical results for neutral quark matter are summarized in 
Figs.~\ref{phasediagram} and \ref{phasediagram-strong}. These are 
the phase diagrams in the plane of temperature and quark chemical 
potential, obtained in the mean-field approximation in model 
(\ref{Lagrangian}) in the case of an intermediate diquark coupling 
strength, $G_D=\frac34 G_S$, and in the case of a strong coupling,
$G_D=G_S$, respectively. The corresponding dynamical quark masses, 
gap parameters, and three charge chemical potentials are displayed 
in Figs.~\ref{plot0-20-40} and \ref{plot0-40-60-strong}, respectively.
All quantities are plotted as functions of $\mu$ for three different 
fixed values of the temperature: $T=0,20,40$~MeV in the case of 
$G_D=\frac34 G_S$ (see Fig.~\ref{plot0-20-40}) and $T=0,40,60$~MeV 
in the case of $G_D=G_S$ (see Fig.~\ref{plot0-40-60-strong}). 

Let us begin with the results in the case of the diquark coupling being
$G_D=\frac34 G_S$. In the region of small quark chemical potentials 
and low temperatures, the phase diagram is dominated by the normal 
phase in which the approximate chiral symmetry is broken, and in 
which quarks have relatively large constituent masses. This is 
denoted by $\chi$SB in Fig.~\ref{phasediagram}. With increasing 
the temperature, this phase changes smoothly into the NQ phase 
in which quark masses are relatively small. Because of explicit 
breaking of the chiral symmetry in the model at hand, there 
is no need for a phase transition between the two regimes. 

However, as pointed out above, the symmetry argument does not 
exclude the possibility of an ``accidental'' (first-order) chiral 
phase transition. As expected, at lower temperatures we find a line 
of first-order chiral phase transitions. It is located within a 
relatively narrow window of the quark chemical potentials 
($336~\mbox{MeV} \lesssim \mu \lesssim 368~\mbox{MeV}$) which are 
of the order of the vacuum values of the light-quark constituent 
masses. (For the parameters used in our calculations one obtains 
$M_u=M_d=367.7\mbox{~MeV}$ and $M_s=549.5\mbox{~MeV}$ in vacuum 
\cite{RKH}.) At this critical line, the quark chiral condensates, 
as well as the quark constituent masses, change discontinuously. 
With increasing temperature, the size of the discontinuity decreases, 
and the line terminates at the endpoint located at $(T_{\rm cr},
\mu_{\rm cr})\approx (56,336)~\mbox{MeV}$, see Fig.~\ref{phasediagram}. 

The location of the critical endpoint is consistent with other mean 
field studies of NJL models with similar sets of parameters 
\cite{cr-pt-NJL,BO,MB}. This agreement does not need to be exact 
because, in contrast to the studies in Refs.~\cite{cr-pt-NJL,BO,MB}, 
here we imposed the condition of electric charge neutrality in quark 
matter. (Note that the color neutrality is satisfied automatically in 
the normal phase.) One may argue, however, that the additional 
constraint of neutrality is unlikely to play a big role in the 
vicinity of the endpoint. 

It is appropriate to mention here that the location of the critical 
endpoint might be affected very much by fluctuations of the composite 
chiral fields. These are not included in the mean-field studies of
the NJL model. In fact, this is probably the main reason for their 
inability to pin down the location of the critical endpoint consistent,
for example, with lattice calculations \cite{lattice}. (It is 
fair to mention that the current lattice calculations are not very 
reliable at nonzero $\mu$ either.) Therefore, the predictions of this 
study, as well as of those in Refs.~\cite{cr-pt-NJL,BO,MB}, regarding 
the critical endpoint cannot be considered as very reliable. 

When the quark chemical potential exceeds some critical value and the 
temperature is not too large, a Cooper instability with respect to 
diquark condensation should develop in the system. Without enforcing 
neutrality, i.e., if the chemical potentials of up and down quarks are 
equal, this happens immediately after the chiral phase transition 
when the density becomes nonzero~\cite{BO}. In the present model, 
this is not the case at low temperatures.

\begin{figure*}[h!]
%%%% h - here, t - top, b - bottom, p - page, ! - as soon as possible 
  \begin{center}
    \includegraphics[width=0.75\textwidth]{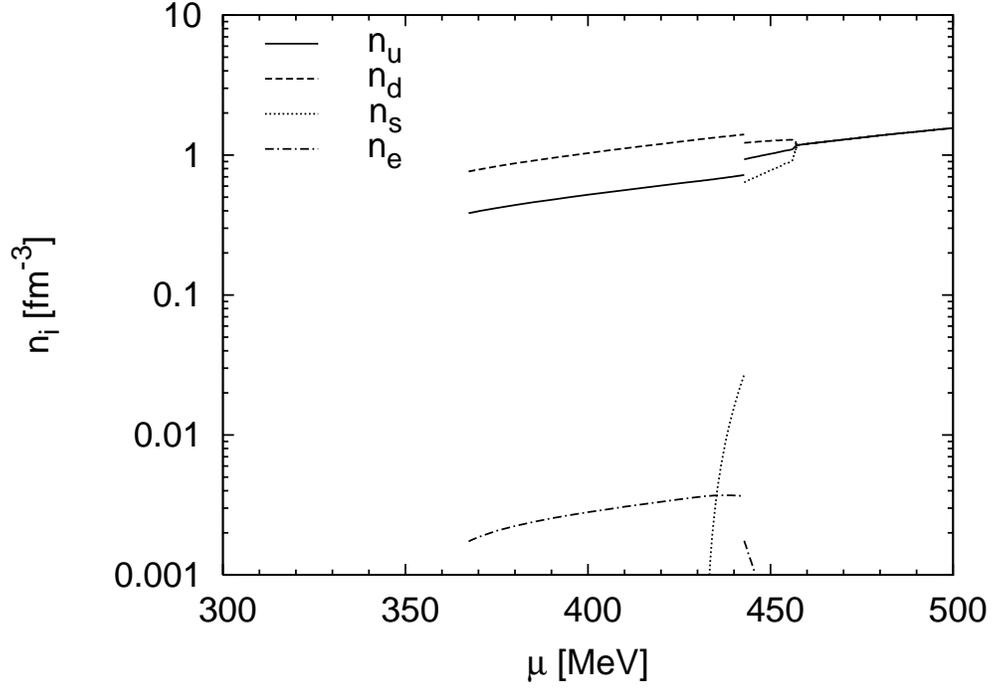}
    \caption{The dependence of the number densities of quarks and
     electrons on the quark chemical potential at $T=0~\mbox{MeV}$
     for diquark coupling strength $G_D=\frac34 G_S$. Note that the 
     densities of all three quark flavors coincide above 
     $\mu=457~\mbox{MeV}$. The density of muons vanishes for
     all values of $\mu$.}
    \label{densT0log}
  \end{center}
\end{figure*}

In order to understand this, let us inspect the various quantities 
at $T=0$ which are displayed in the upper three panels of 
Fig.~\ref{plot0-20-40}. At the chiral phase boundary, the up and 
down quark masses become relatively small, whereas the strange 
quark mass experiences only a moderate drop of about $84$~MeV 
induced by the 't Hooft interaction. This is not sufficient to 
populate any strange quark states at the given chemical potential, 
and the system mainly consists of up and down quarks together with 
a small fraction of electrons, see Fig.~\ref{densT0log}. The electric 
charge chemical potential which is needed to maintain neutrality in 
this regime is between about $-73$ and $-94$~MeV. It turns out that 
the resulting splitting of the up and down quark Fermi momenta is too 
large for the given diquark coupling strength to enable diquark pairing 
and the system stays in the normal phase. 

At $\mu\approx 432~\mbox{MeV}$, the chemical potential felt by the 
strange quarks, $\mu - \mu_Q/3$, reaches the strange quark mass and 
the density of strange quarks becomes nonzero. At first, this density 
is too small to play a sizeable role in neutralizing matter, or in 
enabling strange-nonstrange cross-flavor diquark pairing, see 
Fig.~\ref{densT0log}. The NQ 
phase becomes metastable against the gapless CFL (gCFL) phase at 
$\mu_{\rm gCFL} \approx 443~\mbox{MeV}$. This is the point of a 
first-order phase transition. It is marked by a drop of the 
strange quark mass by about $121$~MeV. As a consequence, strange 
quarks become more abundant and pairing gets easier. Yet, in the 
gCFL phase, the strange quark mass is still relatively large, and 
the standard BCS pairing between strange and light (i.e., up and 
down) quarks is not possible. In contrast to the regular CFL phase, 
the gCFL phase requires a nonzero density of electrons to stay 
electrically neutral. At $T=0$, therefore, one could use the value 
of the electron density as a formal order parameter that distinguishes 
these two phases \cite{gCFL}. 

\begin{figure*}[h!]
%%%% h - here, t - top, b - bottom, p - page, ! - as soon as possible 
  \begin{center}
    \includegraphics[width=0.75\textwidth]{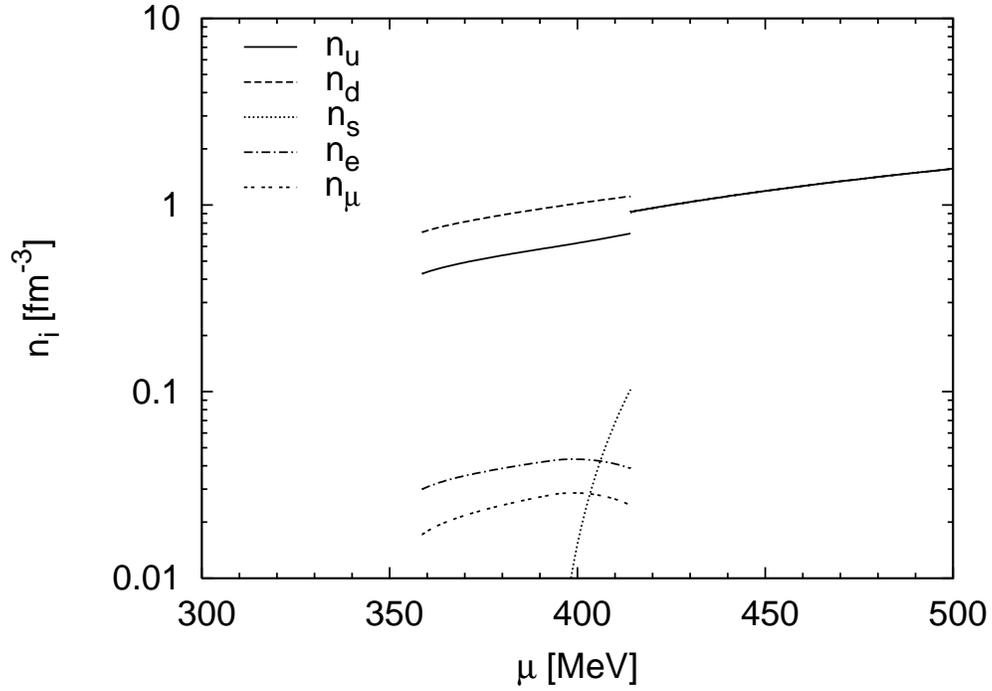}
    \caption{The dependence of the number densities of quarks,
     electrons and muons on the quark chemical potential at 
     $T=0~\mbox{MeV}$ for diquark coupling strength $G_D=G_S$. 
     Note that the densities of all three quark flavors coincide 
     above $\mu=414~\mbox{MeV}$.}
    \label{densT0log_strong}
  \end{center}
\end{figure*}

With increasing the chemical potential further (still at $T=0$), 
the strange quark mass decreases and the cross-flavor Cooper pairing 
gets stronger. Thus, the gCFL phase eventually turns into the regular 
CFL phase at $\mu_{\rm CFL} \approx 457~\mbox{MeV}$. The electron 
density goes to zero at this point, as it should. This is indicated 
by the vanishing value of $\mu_Q$ in the CFL phase, see the upper 
right panel in Fig.~\ref{plot0-20-40}. We remind that the CFL 
phase is neutral because of having equal number densities of all 
three quark flavors, $n_u=n_d=n_s$, see Fig.~\ref{densT0log} and 
\ref{densT0log_strong}. This equality is enforced by the pairing 
mechanism, and this is true even when the quark masses are not 
exactly equal \cite{enforced}. 

Let us mention here that the same NJL model at zero temperature was 
studied previously in Ref.~\cite{SRP}. Our results agree qualitatively 
with those of Ref.~\cite{SRP} only when the quark chemical potential 
is larger than the critical value for the transition to the CFL phase
at $457\mbox{~MeV}$. The appearance of the gCFL phase for $443\lesssim 
\mu \lesssim 457\mbox{~MeV}$ was not recognized in Ref.~\cite{SRP}, 
however. Instead, it was suggested that there exists a narrow (about 
$12\mbox{~MeV}$ wide) window of values of the quark chemical potential 
around $\mu\approx 450\mbox{~MeV}$ in which the 2SC phase is the ground 
state. By carefully checking the same region, we find that the 2SC 
phase does not appear there.

\begin{figure*}[h!]
%%%% h - here, t - top, b - bottom, p - page, ! - as soon as possible 
  \begin{center}
    \includegraphics[width=0.8\textwidth]{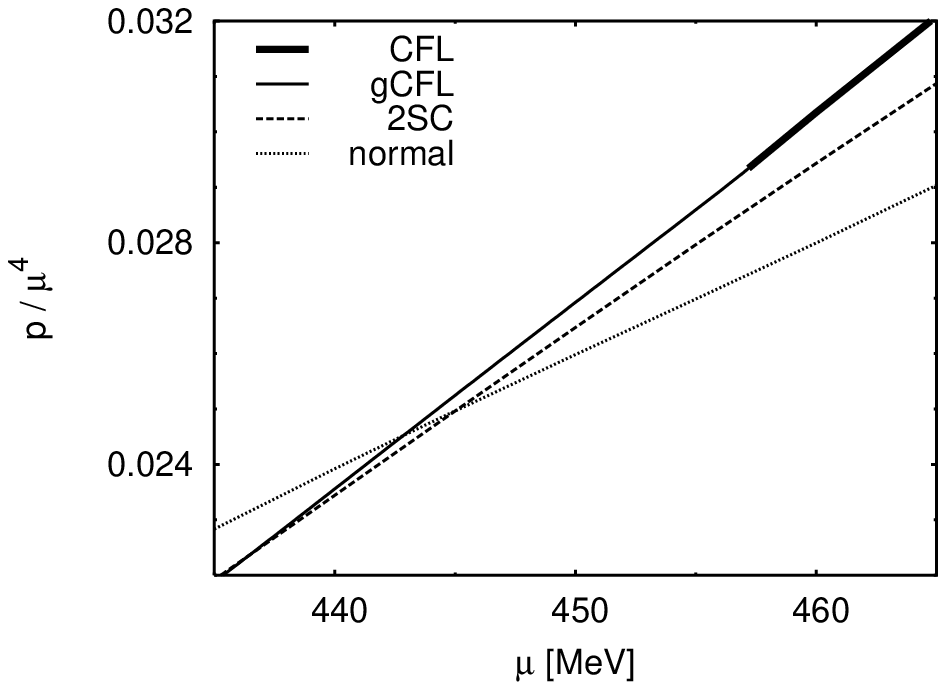}\\
    \caption{Pressure divided by $\mu^4$ for different neutral solutions
             of the gap equations at $T = 0$ as functions of the
             quark chemical potential $\mu$: regular CFL 
             (bold solid line), gapless CFL (thin solid line),
             2SC (dashed line), normal (dotted line). The 
             diquark coupling strength is $G_D=\frac34 G_S$.}
    \label{pmu0}
  \end{center}
\end{figure*}

This is illustrated in Fig.~\ref{pmu0} where the pressure of three 
different solutions is displayed. Had we ignored the gCFL solution 
(thin solid line), the 2SC solution (dashed line) would indeed be the 
most favored one in the interval between $\mu \approx 445$~MeV and 
$\mu \approx 457$~MeV. After including the gCFL phase in the analysis, 
this is no longer the case.

Now let us turn to the case of nonzero temperature. One might suggest 
that this should be analogous to the zero temperature case, except that
Cooper pairing is somewhat suppressed by thermal effects. In contrast 
to this naive expectation, the thermal distributions of quasiparticles
together with the local neutrality conditions open qualitatively new 
possibilities that were absent at $T=0$. As in the case of the two-flavor 
model of Ref.~\cite{g2SC}, a moderate thermal smearing of mismatched 
Fermi surfaces could increase the probability of creating zero-momentum 
Cooper pairs without running into a conflict with Pauli blocking.
This leads to the appearance of several stable color-superconducting 
phases that could not exist at zero temperatures.

With increasing the temperature, the first qualitatively new feature 
in the phase diagram appears when $5 \lesssim T \lesssim 
10~\mbox{MeV}$. In this temperature interval, the NQ phase is 
replaced by the uSC phase when the quark chemical potential 
exceeds the critical value of about $444\mbox{~MeV}$. The 
corresponding transition is a first-order phase transition, see 
Fig.~\ref{phasediagram}. Increasing the chemical potential 
further by several MeV, the uSC phase is then replaced by the gCFL 
phase, and the gCFL phase later turns gradually into the (m)CFL 
phase. (In this study, we do not distinguish between the CFL phase 
and the mCFL phase \cite{phase-d}.) Note that, in the model at hand, 
the transition between the uSC and the gCFL phase is of second order 
in the following two temperature intervals: $5 \lesssim T \lesssim 
9~\mbox{MeV}$ and $T \gtrsim 24~\mbox{MeV}$. On the other hand, it 
is a first-order transition when $9 \lesssim  T \lesssim 24~\mbox{MeV}$. 
Leaving aside its unusual appearance, this is likely to be an ``accidental" 
property in the model for a given set of parameters. For a larger value 
of the diquark coupling, in particular, such a feature does 
not appear, see Fig.~\ref{phasediagram-strong}.

The transition from the gCFL to the CFL phase is a smooth crossover at 
all $T\neq 0$ \cite{phase-d,phase-d1}. The reason is that the electron 
density is not a good order parameter that could be used to distinguish 
the gCFL from the CFL phase when the temperature is nonzero. This is 
also confirmed by our numerical results for the electric charge chemical 
potential $\mu_Q$ in Fig.~\ref{plot0-20-40}. While at zero temperature 
the value of $\mu_Q$ vanishes identically in the CFL phase, this is not 
the case at finite temperatures. 

Another new feature in the phase diagram appears when the temperature 
is above about $11~\mbox{MeV}$. In this case, with increasing the quark
chemical potential, the Cooper instability happens immediately after 
the $\chi$SB phase. The corresponding critical value of the quark 
chemical potential is rather low, about $365~\mbox{MeV}$. The first 
color superconducting phase is the gapless 2SC (g2SC) phase \cite{g2SC}. 
This phase is replaced with the 2SC phase in a crossover transition 
only when $\mu\agt 445~\mbox{MeV}$. The 2SC is then followed by the 
gapless uSC (guSC) phase, by the uSC phase, by the gCFL phase 
and, eventually, by the CFL phase (see Fig.~\ref{phasediagram}). 

In the NJL model at hand, determined by the parameters in 
Eq.~(\ref{model-parameters}), we do not find the dSC phase as the 
ground state anywhere in the phase diagram. This is 
similar to the conclusion of Refs.~\cite{phase-d,phase-d2}, but 
differs from that of Refs.~\cite{dSC,phase-d1}. This should not 
be surprising because, as was noted earlier \cite{phase-d2}, the 
appearence of the dSC phase is rather sensitive to a specific 
choice of parameters in the NJL model.

The phase diagram in Fig.~\ref{phasediagram} has a very specific 
ordering of quark phases. One might ask if this ordering is robust
against the modification of the parameters of the model at hand. Below
we argue that some features are indeed quite robust, while others 
are not. 

It should be clear that the appearance of color-superconducting
phases under the stress of neutrality constraints is very sensitive 
to the strength of diquark coupling. In the case of two-flavor quark 
matter, this was demonstrated very clearly in Ref.~\cite{g2SC} at zero
as well as at nonzero temperatures. Similar conclusions were also 
reached in the study of three-flavor quark matter at zero temperature 
\cite{kyoto}. 

In the model at hand, it is instructive to study the phase diagram in 
the regime of strong diquark coupling, $G_{D}=G_{S}$. The corresponding 
results are summarized in the diagram in Fig.~\ref{phasediagram-strong}.
As we see, the main qualitative difference between the diagrams in 
Figs.~\ref{phasediagram} and \ref{phasediagram-strong} occurs at 
intermediate values of the quark chemical potential. While at $G_{D}
=\frac34 G_{S}$, there is a large region of the g2SC phase sandwiched 
between the low-temperature and high-temperature NQ phases, this is 
not the case at stronger coupling, $G_{D}=G_{S}$. 

The last observation can easily be explained by the fact that with
increasing diquark coupling strength, the condensation energy also
increases and therefore Cooper pairing is favorable, even if there
is a larger mismatch of the Fermi surfaces due to charge neutrality
constraints. Moreover, in the presence of large gaps, the Fermi surfaces 
are smeared over a region of order $\Delta$. Therefore additional thermal 
smearing is of no further help, and it is not surprising that the
thermal effects in a model with sufficiently strong coupling are qualitatively
the same as in models without neutrality constraints imposed: thermal 
fluctuations can only destroy the pairing. In the model with a 
not very strong coupling, on the other hand, the interplay of the 
charge neutrality and thermal fluctuations is more subtle. The normal 
phase of cold quark matter develops a Cooper instability and 
becomes a color superconductor only after a moderate thermal smearing 
of the quark Fermi surfaces is introduced \cite{g2SC}.

Other than this, the qualitative features of the phase diagrams in 
Figs.~\ref{phasediagram} and \ref{phasediagram-strong} are similar. Of 
course, in the case of the stronger coupling, the critical lines lie 
systematically at higher values of the temperature and at lower values 
of the quark chemical potential. In this context one should note that 
the first-order phase boundary between the two normal regimes ``$\chi$SB'' 
and ``NQ'' is insensitive to the diquark coupling. Therefore, upon 
increasing $G_D$ it stays at its place until it is eventually ``eaten'' 
up by the expanding 2SC phase. As a result, there is no longer a critical 
endpoint in Fig.~\ref{phasediagram-strong}, but only a critical
point where the first-order normal($\chi$SB)-2SC phase boundary changes
into second order.

%%%%%%%%%%%%%%%%%%%%%%%%%%%%%%%%%%%%%%%%%%%%%%%%%%%%
%%%%%%%%%%%%%%%%%%%%%%%%%%%%%%%%%%%%%%%%%%%%%%%%%%%%
\section{Conclusions}
\label{conclusions}

In this paper, we studied the $T$--$\mu$ phase diagram of neutral 
three-flavor quark matter within the NJL model of Ref.~\cite{RKH}
in which the chiral symmetry is broken explicitly by small but 
nonzero current quark masses. As in the previous studies 
\cite{phase-d,phase-d1,phase-d2}, we use the mean-field approximation 
in the analysis. In contrast to Refs.~\cite{phase-d,phase-d1,phase-d2}, 
in this paper the constituent quark masses are treated self-consistently 
as dynamically generated quantities. The main results are summarized 
in Figs.~\ref{phasediagram} and \ref{phasediagram-strong}.

By comparing our results with those in Ref.~\cite{phase-d} (see Fig.~11b
there), we notice several important differences. First of all, we observe 
that a self-consistent treatment of quark masses strongly influences the 
competition between different quark phases. As was noticed earlier in
Ref.~\cite{BO}, there exists a subtle interplay between the two main 
effects. On the one hand, the actual values of the quark masses directly 
influence the competition between different normal and color-superconducting 
phases. On the other hand, competing phases themselves determine the magnitude 
of the masses. Very often, this leads to first-order phase transitions,
in which certain regions in the mass-parameter space become inaccessible.

Some differences to the results in Ref.~\cite{phase-d} are related to
a different choice of model parameters. Most importantly, the 
value of the diquark coupling $G_D=\frac34 G_S$ is considerably weaker 
than in the NJL model of Ref.~\cite{phase-d}. This can be easily seen 
by comparing the magnitude of the zero-temperature gap at a given value 
of the quark chemical potential, say at $\mu=500~\mbox{MeV}$, in the two 
models. It is $\Delta_0^{(500)} \approx 76~\mbox{MeV}$ in this paper, and 
it is $\Delta_0^{(500)}\approx 140~\mbox{MeV}$ in Ref.~\cite{phase-d}. 
(Note that the strength of the diquark pairing in 
Ref.~\cite{phase-d1} is even weaker, corresponding to $\Delta_0^{(500)} 
\approx 20~\mbox{MeV}$.) It should be noted that even the case of the 
strong coupling, $G_D=G_S$, which corresponds to $\Delta_0^{(500)} 
\approx 120~\mbox{MeV}$, is still slightly weaker than that in 
Ref.~\cite{phase-d}. In this case, however, the corresponding results 
differ mostly because the quark masses are treated very differently.

Because of the weaker diquark coupling strength, the Cooper instabilities 
in Fig.~\ref{phasediagram} happen systematically at higher values of the 
quark chemical potential than in Ref.~\cite{phase-d}. In particular, 
this is most clearly seen from the critical lines of the transition to 
the (g)CFL phase. Another consequence of the weaker interaction is the 
possibility of a thermal enhancement of the (g)2SC Cooper pairing at 
intermediate values of the quark chemical potential. This kind of 
enhancement was studied in detail in Ref.~\cite{g2SC}. Making use 
of the same arguments, one can tell immediately how the phase diagram 
in Fig.~\ref{phasediagram} should change with increasing or decreasing 
the diquark coupling strength.

In particular, with increasing (decreasing) the diquark coupling 
strength, the region of the (g)2SC phase at intermediate values 
of the quark chemical potential should expand (shrink) along the 
temperature direction. The regions covered by the other (i.e., uSC 
and CFL) phases should have qualitatively the same shape, but shift 
to lower (higher) values of the quark chemical potential and to 
higher (lower) values of the temperature. In the case of strong 
coupling, in particular, these general arguments are confirmed 
by our numerical calculations. The corresponding phase diagram 
is shown in Fig.~\ref{phasediagram-strong}. 

Several comments are in order regarding the choice of the NJL 
model used here. The model is defined by the set of parameters 
in Eq.~(\ref{model-parameters}) which were fitted to reproduce 
several important QCD properties in vacuum \cite{RKH}. (Note
that the same model was also used in Ref.~\cite{SRP}.) It is 
expected, therefore, that this is a reasonable effective model 
of QCD that captures the main features of both chiral and 
color-superconducting pairing dynamics. Also, a relatively small 
value of the cutoff parameter in the model, see Eq.~(\ref{Lambda}), 
should not necessarily be viewed as a bad feature of the model. 
In fact, this might simply mimic a natural property of the full 
theory in which the coupling strength of relevant interactions is 
quenched at large momenta. 

In this relation, note that the approach of Ref.~\cite{phase-d1} 
regarding the cutoff parameter in the NJL model is very different. 
It is said there that a large value of this parameter is beneficial 
in order to extract results which are insensitive to a specific 
choice of the cutoff. However, we do not find any physical argument 
that would support this requirement. Instead, we insist on having 
an effective model that describes reasonably well the QCD properties 
at zero quark chemical potential. We do not pretend, of course, that 
a naive extrapolation of the model to large densities can be 
rigorously justified. In absence of a better alternative, however, 
this seems to be the only sensible choice. 

The results of this paper might be relevant for understanding 
the physics of (hybrid) neutron stars with quark cores, in which 
the deleptonization is completed. In order to obtain a phase 
diagram that could be applied to protoneutron stars, one has to 
generalize the analysis to take into account neutrino trapping.
This work is in progress now \cite{pd-nu}.

In the end, it might be appropriate to mention that, despite the 
progress in our understanding of the phase diagram of neutral 
dense quark matter, there still exists a fundamental problem here. 
The reason is that some regions of the phase diagrams in 
Figs.~\ref{phasediagram} and \ref{phasediagram-strong} correspond 
to phases that are known to be unstable \cite{instability}. 
Of course, it is of prime importance to resolve this issue.

{\em Note added.} While writing our paper, we learned that 
a partially overlapping study is being done by D.~Blaschke,
S.~Fredriksson, H.~Grigorian, A.M.~\"{O}zta\c{s}, and F.~Sandin 
\cite{BFGOS}.

%%%%%%%%%%%%%%%%%%%%%%%%%%%%%%%%%%%%%%%%%%%%%%%%%%%%
%%%%%%%%%%%%%%%%%%%%%%%%%%%%%%%%%%%%%%%%%%%%%%%%%%%%
\begin{acknowledgments}
This work was supported in part by the Virtual Institute of the 
Helmholtz Association under grant No. VH-VI-041 and by Gesellschaft 
f\"{u}r Schwerionenforschung (GSI) and by Bundesministerium f\"{u}r 
Bildung und Forschung (BMBF). S.~R. thanks for using the Center 
for Scientific Computing (CSC) of the Johann Wolfgang 
Goethe-Universit\"at Frankfurt am Main.
\end{acknowledgments}

\medskip

\appendix

\section{Evaluation of the thermodynamic potential}
\label{tpot}

We use the 
following ordering of the quark field components:
\begin{equation}
\psi =
\left(
\psi_u^r,
\psi_d^r,
\psi_s^r,
\psi_u^g,
\psi_d^g,
\psi_s^g,
\psi_u^b,
\psi_d^b,
\psi_s^b
\right)^T \; .
\end{equation}
In this basis, the matrices of quark current and constituent masses 
read
\begin{equation}
\hat{m} = \diag \left( m_u, m_d, m_s, m_u, m_d, m_s, m_u, m_d,
m_s \right)
\end{equation}
and
\begin{equation}
\hat{M} = \diag \left(M_u, M_d, M_s, M_u, M_d, M_s, M_u, M_d,
M_s \right)\; ,
\label{hat-M} 
\end{equation}
respectively, with $M_\alpha$ given by Eq.~(\ref{Mi}). 
Moreover, the matrix of quark chemical potentials takes the general
form
\begin{equation}
\hat\mu = \diag \left( \mu_u^r, \mu_d^r, \mu_s^r, \mu_u^g,
\mu_d^g, \mu_s^g ,\mu_u^b, \mu_d^b, \mu_s^b \right)
\label{hat-mu} \; .
\end{equation}
Finally, the explicit color-flavor structure of the gap matrices $\Phi^\pm$
[see Eq.~(\ref{Phim})] is given by
\begin{equation}
\label{Phi}
\Phi^- = - \gamma_5 
\left( 
\begin{array}{@{\extracolsep{-3.2mm}}ccccccccc}
0 & 0 & 0 & 0 & \phantom{-} \Delta_3 & 0 & 0 & 0 & \phantom{-} \Delta_2 \\
0 & 0 & 0 & - \Delta_3 & 0 & 0 & 0 & 0 & 0 \\
0 & 0 & 0 & 0 & 0 & 0 & - \Delta_2 & 0 & 0 \\
0 & - \Delta_3 & 0 & 0 & 0 & 0 & 0 & 0 & 0 \\
\phantom{-} \Delta_3 & 0 & 0 & 0 & 0 & 0 & 0 & 0 & \phantom{-} \Delta_1 \\
0 & 0 & 0 & 0 & 0 & 0 & 0 & - \Delta_1 & 0 \\
0 & 0 & - \Delta_2 & 0 & 0 & 0 & 0 & 0 & 0 \\
0 & 0 & 0 & 0 & 0 & - \Delta_1 & 0 & 0 & 0 \\
\phantom{-} \Delta_2 & 0 & 0 & 0 & \phantom{-} \Delta_1 & 0 & 0 & 0 & 0
\end{array}
\right)
\; ,
\end{equation}
and $\Phi^+ = - \left(\Phi^-\right)^\dagger$. By making use of the 
symmetries in the model, we choose all three parameters $\Delta_c$ to 
be real.

In order to calculate the last term in the thermodynamical potential 
in Eq.~(\ref{Omega}), it is useful to rewrite the determinant of 
the inverse full propagator as 
    $\det S^{-1} = \det \left(\gamma_0 \gamma_0 S^{-1} \right) 
         = \det (\gamma_0) \det \left( \gamma_0 S^{-1} \right) 
                         = \det \left( \gamma_0 S^{-1} \right)$. 
One should also note that the matrices $\gamma_0 [G_0^\pm]^{-1}$ 
and $\gamma_0 \Phi^\pm$ can be expressed in terms of the spin 
projectors,
\begin{eqnarray}
\gamma_0 [G_0^\pm]^{-1} &=&
\sum_s
\left(
\begin{array}{cc}
k_0 \pm \hat{\mu} - \hat{M} & -sk \\
-sk & k_0 \pm \hat{\mu} + \hat{M}
\end{array}
\right)
\mathcal{P}_s \; , \quad
\label{G0-Ps} \\
%%%%%%%%%%%%%%%%%%%%%%%%%%%
\gamma_0 \Phi^\pm &=&
\pm \sum_s
\left(
\begin{array}{cc}
\ \ \, 0 \ & \hat{\Phi}  \\
-\hat{\Phi} \ & 0
\end{array}
\right)
\mathcal{P}_s \; ,
\label{Phi-Ps}
\end{eqnarray}
where the two projectors are defined as
\begin{equation}
\mathcal{P}_s = \frac{1}{2} ( 1 + s \, \fett{\sigma} \cdot
\hat{\fettu{k}} ) \; , \quad \mbox{for}\quad  \quad s=\pm \; .
\label{P_s}
\end{equation}
Here $\hat{\fettu{k}}\equiv \fettu{k}/k$, $k\equiv |\fettu{k}|$, 
and $\hat{\Phi}$ represents only the color-flavor part of the 
gap matrix in Eq.~(\ref{Phi}), i.e., $\Phi^\pm \equiv \pm \gamma_5 
\hat{\Phi}$. By making use of the definition in Eq.~(\ref{off-d}), 
as well as Eqs.~(\ref{G0-Ps}) and (\ref{Phi-Ps}), we obtain the 
following representation:
\begin{equation}
\gamma_0 S^{-1} = \sum_s \hat{S}_s^{-1}
\mathcal{P}_s \; ,
\end{equation}
where $\hat{S}_s^{-1} = \openone k_0 - \mathcal{M}_s$, and 
\begin{equation}
\mathcal{M}_s =
\left(
\begin{array}{cccc}
- \hat{\mu} + \hat{M} & sk & 0 & \hat{\Phi} \\
sk & - \hat{\mu} - \hat{M} & -\hat{\Phi} & 0 \\
0 & -\hat{\Phi} & \hat{\mu} + \hat{M} & sk \\
\hat{\Phi} & 0 & sk & \hat{\mu} - \hat{M}
\end{array}
\right),
\label{matSs}
\end{equation}
(with $s=\pm$) is real and symmetric. Since there is no explicit 
energy dependence in $\mathcal{M}_s$, their eigenvalues $\epsilon_{i}$ 
determine the quasiparticle dispersion relations, $k_0=\epsilon_{i}(k)$. 
By using the matrix relation $\ln \det (A) = \Tr \ln (A)$ as well as the 
properties of projectors $\mathcal{P}_s$, we derive
\begin{equation}
\ln \det \left( \gamma_0 S^{-1} \right)
= \ln \left( \det \hat{S}_+^{-1} \cdot \det
\hat{S}_-^{-1} \right) \; .
\label{lndet}
\end{equation}
It turns out that the two determinants appearing on the right hand side 
of this equation are equal, i.e., $\det\hat{S}_-^{-1}=\det\hat{S}_+^{-1}$. 
{From} the physics viewpoint, this identity reflects the twofold spin degeneracy 
of the spectrum of quark quasiparticles. The formal proof of this 
degeneracy is straightforward after noticing that the following matrix 
relation is satisfied:
\begin{equation}
\hat{S}_{-s}^{-1} = \mathcal{R}
\hat{S}_s^{-1} \mathcal{R}^{-1} \; ,
\end{equation}
where
\begin{equation}
\mathcal{R}=
\left(
\begin{array}{rrrr}
1 & 0 & 0 & \ \ 0 \\
0 & -1 & 0 & \ \ 0 \\
0 & 0 & -1 & \ \ 0 \\
0 & 0 & 0 & \ \ 1
\end{array}
\right)
\end{equation}
is a unitary matrix with unit determinant, $\det\mathcal{R}=1$.

Another observation, which turns out to be helpful in the calculation, 
is that the determinant $\det\hat{S}_{s}^{-1}(k_0)$ is an even function 
of $k_0$, i.e., $\det\hat{S}_{s}^{-1}(-k_0)=\det\hat{S}_{s}^{-1}(k_0)$. 
This is a formal consequence of the following matrix relation:
\begin{equation}
\hat{S}_s^{-1} \left( - k_0 \right) = -\mathcal{B}
\hat{S}_s^{-1} \left( k_0 \right) \mathcal{B}^{-1} \; ,
\end{equation}
where the explicit form of the unitary matrix $\mathcal{B}$ is
\begin{equation}
\mathcal{B} =
\left(
\begin{array}{rrrr}
0 & \ \ 0 & 0 & \ \ i \\
0 & \ \ 0 & -i & \ \ 0 \\
0 & \ \ i & 0 & \ \ 0 \\
-i & \ \ 0 & 0 & \ \ 0
\end{array}
\right).
\end{equation}
It satisfies $\det\mathcal{B}=1$.
The invariance of the determinant $\det\hat{S}_{s}^{-1}(k_0)$ with 
respect to the change of the energy sign, $k_0\to -k_0$, is directly 
related to the use of the Nambu-Gorkov basis for quark fields. In 
this basis, for each quasiparticle excitation with a positive 
energy $k_0=\epsilon(k)$, there exists a corresponding excitation 
with a negative energy $k_0=-\epsilon(k)$. Therefore, the result 
for the determinant should read
\begin{equation}
\det\left(S^{-1}\right) 
= \prod_{i=1}^{18}\left(k_0^2-\epsilon_i^2\right)^2.
\end{equation}
In order to simplify the numerical calculation of the eigenvalues of the 
matrix $\mathcal{M}_{+}$, defined in Eq.~(\ref{matSs}), we first 
write it in a block-diagonal form. The total dimension of this matrix 
is $36 \times 36$. With a proper ordering of its rows and columns, 
it decomposes into $6$ diagonal blocks of dimension $4 \times 4$ and one 
diagonal block of dimension $12 \times 12$. The explicit form of 
these blocks
reads
%\begin{widetext}
\begin{subequations} 
%%%%%%%%%%%%%%%%%%%%%%%%%%%%%%%%%%%%%%%%
\begin{equation}
\mathcal{M}_{+}^{(1)}=  
\left( 
\begin{array}{cccc}
 - \mu_d^r + M_d & k & 0 & - \Delta_3 \\
k &  - \mu_d^r - M_d & \Delta_3 & 0 \\
0 & \Delta_3 & \mu_u^g + M_u & k \\
- \Delta_3 & 0 & k & \mu_u^g - M_u
\end{array}
\right) \; ,
\end{equation}
%%%%%%%%%%%%%%%%%%%%%%%%%%%%%%%%%%%%%%%%
\begin{equation}
\mathcal{M}_{+}^{(2)}=  
\left( 
\begin{array}{cccc}
 \mu_d^r - M_d & k & 0 & - \Delta_3 \\
k &  \mu_d^r + M_d & \Delta_3 & 0 \\
0 & \Delta_3 & -\mu_u^g - M_u & k \\
- \Delta_3 & 0 & k & -\mu_u^g + M_u
\end{array}
\right) \; ,
\end{equation}
%%%%%%%%%%%%%%%%%%%%%%%%%%%%%%%%%%%%%%%%
\begin{equation}
\mathcal{M}_{+}^{(3)}=  
\left( 
\begin{array}{cccc}
 - \mu_s^r + M_s & k & 0 & - \Delta_2 \\
k &  - \mu_s^r - M_s & \Delta_2 & 0 \\
0 & \Delta_2 & \mu_u^b + M_u & k \\
- \Delta_2 & 0 & k & \mu_u^b - M_u
\end{array}
\right) \; ,
\end{equation}
%%%%%%%%%%%%%%%%%%%%%%%%%%%%%%%%%%%%%%%%
\begin{equation}
\mathcal{M}_{+}^{(4)}=  
\left( 
\begin{array}{cccc}
 \mu_s^r - M_s & k & 0 & - \Delta_2 \\
k &  \mu_s^r + M_s & \Delta_2 & 0 \\
0 & \Delta_2 & -\mu_u^b - M_u & k \\
- \Delta_2 & 0 & k & -\mu_u^b + M_u
\end{array}
\right) \; ,
\end{equation}
%%%%%%%%%%%%%%%%%%%%%%%%%%%%%%%%%%%%%%%%
\begin{equation}
\mathcal{M}_{+}^{(5)}=  
\left( 
\begin{array}{cccc}
 - \mu_s^g + M_s & k & 0 & - \Delta_1 \\
k &  - \mu_s^g - M_s & \Delta_1 & 0 \\
0 & \Delta_1 & \mu_d^b + M_d & k \\
- \Delta_1 & 0 & k & \mu_d^b - M_d
\end{array}
\right) \; ,
\end{equation}
%%%%%%%%%%%%%%%%%%%%%%%%%%%%%%%%%%%%%%%%
\begin{equation}
\mathcal{M}_{+}^{(6)}=  
\left( 
\begin{array}{cccc}
 \mu_s^g - M_s & k & 0 & - \Delta_1 \\
k &  \mu_s^g + M_s & \Delta_1 & 0 \\
0 & \Delta_1 & -\mu_d^b - M_d & k \\
- \Delta_1 & 0 & k & -\mu_d^b + M_d
\end{array}
\right) \; ,
\end{equation}
%%%%%%%%%%%%%%%%%%%%%%%%%%%%%%%%%%%%%%%%
and
%%%%%%%%%%%%%%%%%%%%%%%%%%%%%%%%%%%%%%%%
\begin{widetext}
\begin{equation}
\mathcal{M}_{+}^{(7)} =
\left(
\begin{array}{@{\extracolsep{-3.5mm}}cccccccccccc}
%%% 1st row of 4x4 blocks 
-\mu_u^r -M_u &       k      &       0      &      0        &
       0      &       0      &       0      &   -\Delta_3   &
       0      &       0      &       0      &   -\Delta_2   \\
       k      & -\mu_u^r+M_u &       0      &       0       &
       0      &       0      &   \Delta_3   &       0       &
       0      &       0      &   \Delta_2   &       0       \\
       0      &       0      & \mu_u^r -M_u &       k       &
       0      &   \Delta_3   &       0      &       0       &
       0      &   \Delta_2   &       0      &       0       \\
       0      &       0      &       k      & \mu_u^r +M_u  &
   -\Delta_3  &       0      &       0      &       0       &
   -\Delta_2  &       0      &       0      &       0       \\[2mm]
%%% 2nd row of 4x4 blocks 
       0      &       0      &       0      &   -\Delta_3   &
\quad 
-\mu_d^g -M_d &       k      &       0      &      0        &
       0      &       0      &       0      &   -\Delta_1   \\
       0      &       0      &   \Delta_3   &       0       &
       k      & -\mu_d^g+M_d &       0      &       0       &
       0      &       0      &   \Delta_1   &       0       \\
       0      &   \Delta_3   &       0      &       0       &
       0      &       0      & \mu_d^g -M_d &       k       &
       0      &   \Delta_1   &       0      &       0       \\
   -\Delta_3  &       0      &       0      &       0       &
       0      &       0      &       k      & \mu_d^g +M_d  &
   -\Delta_1  &       0      &       0      &       0       \\[2mm]
%%% 3rd row of 4x4 blocks 
       0      &       0      &       0      &   -\Delta_2   &
       0      &       0      &       0      &   -\Delta_1   &
\quad 
-\mu_s^b -M_s &       k      &       0      &      0        \\
       0      &       0      &   \Delta_2   &       0       &
       0      &       0      &   \Delta_1   &       0       &
       k      & -\mu_s^b+M_s &       0      &       0       \\
       0      &   \Delta_2   &       0      &       0       &
       0      &   \Delta_1   &       0      &       0       &
       0      &       0      & \mu_s^b -M_s &       k       \\
   -\Delta_2  &       0      &       0      &       0       &
   -\Delta_1  &       0      &       0      &       0       &
       0      &       0      &       k      & \mu_s^b +M_s 
\end{array}
\right) .
\end{equation}
\end{widetext}
%%%%%%%%%%%%%%%%%%%%%%%%%%%%%%%%%%%%%%%%
\end{subequations}
%\end{widetext}
Out of 36 eigenvalues from all seven blocks, there are 18 positive 
and 18 negative eigenvalues. Out of total 18 positive eigenvalues, 
9 of them correspond to quark type quasiparticles and the other 9 
correspond to antiquark type quasiparticles. In our calculation, we 
extract all 36 eigenvalues numerically and then use them in the 
calculation of the pressure, see Eqs.~(\ref{det-S}) through 
(\ref{pressure}). 

Here, it might be interesting to note that the eigenvalues of the 
$4\times 4$ matrices can be calculated analytically in the limit 
when two quark masses appearing in each of them are equal. For example, 
when $M_d=M_u$, the four eigenvalues of matrix $\mathcal{M}_{+}^{(1)}$ 
are given by
\begin{equation}
\lambda^{(1)}_{i}=\pm \sqrt{\left(\frac{\mu_d^r+\mu_u^g}{2} 
\pm\sqrt{M_u^2+k^2}\right)^2+\Delta_3^2}
-\frac{\mu_d^r-\mu_u^g}{2}\; ,
\end{equation}
while the eigenvalues of $\mathcal{M}_{+}^{(2)}$ differ only by 
the sign in front of the second term,
\begin{equation}
\lambda^{(2)}_{i}=\pm \sqrt{\left(\frac{\mu_d^r+\mu_u^g}{2} 
\pm\sqrt{M_u^2+k^2} \right)^2+\Delta_3^2}
+\frac{\mu_d^r-\mu_u^g}{2}\; .
\end{equation}
When the value of $\delta M\equiv M_d-M_u$ is nonzero but small, 
the corrections to the above eigenvalues are 
$\pm M_u \delta M/(2\sqrt{M_u^2+k^2})$ with the plus sign in the 
case of antiparticle modes, and the minus sign in the case of 
particle modes. The eigenvalues of $\mathcal{M}_{+}^{(3)}$ and 
$\mathcal{M}_{+}^{(4)}$ in the limit $M_s=M_u$, as well as the 
eigenvalues of $\mathcal{M}_{+}^{(5)}$ and $\mathcal{M}_{+}^{(6)}$ 
in the limit $M_s=M_d$, are similar.

%%%%%%% begin the bibliography %%%%%%%%%%%%%%%%%%%%%

%%%%%%%%%%%% end the bibliography %%%%%%%%%%%%%%%%%%%%%%%

%%%%%%%%%%%%%%%%%%%%%%%%%%%%%%%%%%%%%%%%%%%%%%%%%%%%%%%%%%%%%%%%%%%%%
%%%%%%%%%%%%%%%%%%%%%%%%%%%%%%%%%%%%%%%%%%%%%%%%%%%%%%%%%%%%%%%%%%%%%

%%%%%%%%%%%%%%%%%%%%%%%%%%%%%%%%%%%%%%%%%%%%%%%%%%%%%%%%%%%%%%%%%%%%%
%%%%%%%%%%%%%%%%%%%%%%%%%%%%%%%%%%%%%%%%%%%%%%%%%%%%%%%%%%%%%%%%%%%%%

%%%%%%%%%%%%%%%%%%%%%%%%%%%%%%%%%%%%%%%%%%%%%%%%%%%%%%%%%%%%%%%%%%%%%
%%%%%%%%%%%%%%%%%%%%%%%%%%%%%%%%%%%%%%%%%%%%%%%%%%%%%%%%%%%%%%%%%%%%%

%%%%%%%%%%%%%%%%%%%%%%%%%%%%%%%%%%%%%%%%%%%%%%%%%%%%%%%%%%%%%%%%%%%%%
%%%%%%%%%%%%%%%%%%%%%%%%%%%%%%%%%%%%%%%%%%%%%%%%%%%%%%%%%%%%%%%%%%%%%
%%%%%%%%%%%%%%%%%%%%%%%%%%%%%%%%%%%%%%%%%%%%%%%%%%%%%%%%%%%%%%%%%%%%%

%%%%%%%%%%%%%%%%%%%%%%%%%%%%%%%%%%%%%%%%%%%%%%%%%%%%%%%%%%%%%%%%%%%%%
%%%%%%%%%%%%%%%%%%%%%%%%%%%%%%%%%%%%%%%%%%%%%%%%%%%%%%%%%%%%%%%%%%%%%


\begin{thebibliography}{99}

\bibitem{reviews} K.~Rajagopal and F.~Wilczek,
%``The condensed matter physics of QCD,''
hep-ph/0011333;
%%CITATION = HEP-PH 0011333;%%
M.~Alford,
%``Color superconducting quark matter,''
Ann.\ Rev.\ Nucl.\ Part.\ Sci.\  {\bf 51}, 131 (2001);
%%CITATION = HEP-PH 0102047;%%
T.~Sch{\"a}fer,
%``Quark matter,''
hep-ph/0304281;
%%CITATION = HEP-PH 0304281;%%
D.~H.~Rischke, 
%``The quark-gluon plasma in equilibrium,''
Prog.\ Part.\ Nucl.\ Phys.\  {\bf 52}, 197 (2004);
%%CITATION = NUCL-TH 0305030;%%
H.-C.~Ren,
%``Color superconductivity of QCD at high baryon density,''
hep-ph/0404074;
%%CITATION = HEP-PH 0404074;%%
M.~Huang,
%``Color superconductivity at moderate baryon density,''
hep-ph/0409167;
%%CITATION = HEP-PH 0409167;%%
I.~A.~Shovkovy,
%``Two lectures on color superconductivity,''
nucl-th/0410091.
%%CITATION = NUCL-TH 0410091;%%

\bibitem{absence2sc} 
M.~Alford and K.~Rajagopal,
%``Absence of two-flavor color superconductivity in compact stars,''
JHEP {\bf 0206}, 031 (2002).
%%CITATION = HEP-PH 0204001;%%

\bibitem{SRP}
A.~W.~Steiner, S.~Reddy, and M.~Prakash,
%``Color-neutral superconducting quark matter,''
Phys.\ Rev.\ D {\bf 66}, 094007 (2002).
%%CITATION = HEP-PH 0205201;%%

\bibitem{mei} M.~Huang, P.~F.~Zhuang and W.~Q.~Chao,
%``Charge neutrality effects on 2-flavor color superconductivity,''
Phys.\ Rev.\ D {\bf 67}, 065015 (2003).
%%CITATION = HEP-PH 0207008;%%

\bibitem{g2SC} I.~Shovkovy and M.~Huang,
%``Gapless two-flavor color superconductor,''
Phys.\ Lett.\ B {\bf 564}, 205 (2003);
%%CITATION = HEP-PH 0302142;%%
M.~Huang and I.~Shovkovy,
%``Gapless color superconductivity at zero and at finite temperature,''
Nucl.\ Phys.\ A {\bf 729}, 835 (2003).
%%CITATION = HEP-PH 0307273;%%

\bibitem{gCFL} M.~Alford, C.~Kouvaris, and K.~Rajagopal,
%``Gapless color-flavor-locked quark matter,''
Phys.\ Rev.\ Lett.\  {\bf 92}, 222001 (2004);
%%CITATION = HEP-PH 0311286;%%
Phys.\ Rev.\ D {\bf 71}, 034002 (2004).
%%CITATION = HEP-PH 0406137;%%

\bibitem{phase-d} 
S.~B.~R\"{u}ster, I.~A.~Shovkovy and D.~H.~Rischke, 
%``Phase diagram of dense neutral three-flavor quark matter,''
Nucl.\ Phys.\ A {\bf 743}, 127 (2004).
%%CITATION = HEP-PH 0405170;%%

\bibitem{cfl} M.~G.~Alford, K.~Rajagopal, and F.~Wilczek,
%``Color-flavor locking and chiral symmetry breaking in ...''
Nucl.\ Phys.\ {\bf B537}, 443 (1999).
%%CITATION = HEP-PH 9804403;%%

\bibitem{weakCFL} I.~A.~Shovkovy and L.~C.~R.~Wijewardhana,
Phys.\ Lett.\ B {\bf 470}, 189 (1999);
%%CITATION = HEP-PH 9910225;%%
T.~Sch\"{a}fer,
Nucl.\ Phys.\ {\bf B575}, 269 (2000).
%%CITATION = HEP-PH 9909574;%%

\bibitem{dSC} K.~Iida, T.~Matsuura, M.~Tachibana, and T.~Hatsuda,
%``Melting pattern of diquark condensates in quark matter,''
Phys.\ Rev.\ Lett.\  {\bf 93}, 132001 (2004).
%%CITATION = HEP-PH 0312363;%%

\bibitem{cs} M.~Alford, K.~Rajagopal, and F.~Wilczek,
Phys.\ Lett.\ B {\bf 422}, 247 (1998);
%%CITATION = HEP-PH 9711395;%%
R.~Rapp, T.~Sch{\"a}fer, E.~V.~Shuryak, and M.~Velkovsky,
%``Diquark Bose condensates in high density matter and instantons,''
Phys.\ Rev.\ Lett.\  {\bf 81}, 53 (1998).
%%CITATION = HEP-PH 9711396;%%

\bibitem{weak} D.~T.~Son,
Phys.\ Rev.\ D {\bf 59}, 094019 (1999);
%%CITATION = HEP-PH 9812287;%%
T.~Sch\"{a}fer and F.~Wilczek,
Phys.\ Rev.\ D {\bf 60}, 114033 (1999);
%%CITATION = HEP-PH 9906512;%%
D.~K.~Hong, V.~A.~Miransky, I.~A.~Shovkovy, and L.~C.~R.~Wijewardhana,
Phys.\ Rev.\ D {\bf 61}, 056001 (2000);
%%CITATION = HEP-PH 9906478;%%
R.~D.~Pisarski and D.~H.~Rischke,
Phys.\ Rev.\ D {\bf 61}, 051501 (2000);
%%CITATION = NUCL-TH 9907041;%%
W.~E.~Brown, J.~T.~Liu, and H.-C.~Ren,   
Phys.\ Rev.\ D {\bf 61}, 114012 (2000).
%%CITATION = HEP-PH 9908248;%%

\bibitem{phase-d1}
K.~Fukushima, C.~Kouvaris and K.~Rajagopal,
%``Heating (gapless) color-flavor locked quark matter,''
Phys.\ Rev.\ D {\bf 71}, 034002 (2005).
%%CITATION = HEP-PH 0408322;%%

\bibitem{phase-d2}
I.~A.~Shovkovy, S.~B.~R\"{u}ster and D.~H.~Rischke,
%``Gapless phases of color-superconducting matter,''
J.\ Phys.\ G {\bf 31}, S849 (2005).
%%CITATION = NUCL-TH 0411040;%%

\bibitem{kyoto} H.~Abuki, M.~Kitazawa and T.~Kunihiro,
%``How does the dynamical chiral condensation affect the three-flavor neutral
%quark matter?,''
hep-ph/0412382.
%%CITATION = HEP-PH 0412382;%%

\bibitem{mix} F.~Neumann, M.~Buballa and M.~Oertel,
%``Mixed phases of color superconducting quark matter,''
Nucl.\ Phys.\ A {\bf 714}, 481 (2003);
%%CITATION = HEP-PH 0210078;%%
I.~Shovkovy, M.~Hanauske and M.~Huang,
%``Nonstrange hybrid compact stars with color superconducting matter,''
Phys.\ Rev.\ D {\bf 67}, 103004 (2003);
%%CITATION = HEP-PH 0303027;%%
S.~Reddy and G.~Rupak,
%``Phase Structure of 2-Flavor Quark Matter: Heterogeneous Superconductors,''
Phys.\ Rev.\ C {\bf 71}, 025201 (2005).
%%CITATION = NUCL-TH 0405054;%%

\bibitem{cryst} M.~G.~Alford, J.~A.~Bowers and K.~Rajagopal,
%``Crystalline color superconductivity,''
Phys.\ Rev.\ D {\bf 63}, 074016 (2001);
%%CITATION = HEP-PH 0008208;%%
R.~Casalbuoni, R.~Gatto, M.~Mannarelli, and G.~Nardulli,
%``Anisotropy parameters for the effective description of ...''
Phys.\ Rev.\ D {\bf 66}, 014006 (2002);
%%CITATION = HEP-PH 0201059;%%
I.~Giannakis, J.~T.~Liu, and H.-C.~Ren,
%``Angular momentum mixing in crystalline color superconductivity,''
Phys.\ Rev.\ D {\bf 66}, 031501 (2002).
%%CITATION = HEP-PH 0202138;%%

\bibitem{goldstones} M.~Buballa,
%``NJL-model description of Goldstone boson condensation in the color-flavor
%locked phase,''
Phys.\ Lett.\ B {\bf 609}, 57 (2005);
%%CITATION = HEP-PH 0410397;%%
M.~M.~Forbes,
%``Kaon condensation in an NJL model at high density,''
hep-ph/0411001.
%%CITATION = HEP-PH 0411001;%%

\bibitem{RKH} P.~Rehberg, S.~P.~Klevansky and J.~H\"{u}fner,
%``Hadronization in the SU(3) Nambu--Jona-Lasinio model,''
Phys.\ Rev.\ C {\bf 53}, 410 (1996).
%%CITATION = HEP-PH 9506436;%%

\bibitem{RSSV00}
  R.~Rapp, T.~Sch\"afer, E.~V.~Shuryak and M.~Velkovsky,
  %``High-density QCD and instantons,''
  Annals Phys.\  {\bf 280}, 35 (2000).
  %%CITATION = HEP-PH 9904353;%%

\bibitem{MB} M.~Buballa,
%``NJL model analysis of quark matter at large density,''
Phys.\ Rep.\ {\bf 407}, 205 (2005).
%%CITATION = HEP-PH 0402234;%%

\bibitem{Kapusta} J.~I.~Kapusta, {\sl Finite-temperature field
theory}, (University Press, Cambridge, 1989).

\bibitem{pd-nu} S.~B.~R\"uster, V.~Werth, M.~Buballa, 
I.~A.~Shovkovy, and D.~H.~Rischke, in preparation.

\bibitem{cr-pt-NJL} M.~Asakawa and K.~Yazaki,
%``Chiral Restoration At Finite Density And Temperature,''
Nucl.\ Phys.\ A {\bf 504}, 668 (1989);
%%CITATION = NUPHA,A504,668;%%
%``Color superconductivity and chiral symmetry restoration at nonzero  baryon
%density and temperature,''
Nucl.\ Phys.\ B {\bf 538}, 215 (1999);
%%CITATION = HEP-PH 9804233;%%
O.~Scavenius, A.~Mocsy, I.~N.~Mishustin and D.~H.~Rischke,
%``Chiral phase transition within effective models with constituent  quarks,''
Phys.\ Rev.\ C {\bf 64}, 045202 (2001).
%%CITATION = NUCL-TH 0007030;%%

\bibitem{BO}
M.~Buballa and M.~Oertel,
%``Color-flavor unlocking and phase diagram with self-consistently  determined
%strange quark masses,''
Nucl.\ Phys.\ A {\bf 703}, 770 (2002).
%%CITATION = HEP-PH 0109095;%%

\bibitem{lattice} Z.~Fodor and S.~D.~Katz,
%``Critical point of QCD at finite T and mu, lattice results for physical quark
%masses,''
JHEP {\bf 0404}, 050 (2004).
%%CITATION = HEP-LAT 0402006;%%

\bibitem{enforced}
K.~Rajagopal and F.~Wilczek,
%``Enforced electrical neutrality of the color-flavor locked phase,''
Phys.\ Rev.\ Lett.\  {\bf 86}, 3492 (2001).
%%CITATION = HEP-PH 0012039;%%

\bibitem{instability} 
M.~Huang and I.~A.~Shovkovy,
%``Chromomagnetic instability in dense quark matter,''
Phys. Rev. D {\bf 70}, 051501(R) (2004);
%%CITATION = HEP-PH 0407049;%%
%``Screening masses in neutral two-flavor color superconductor,''
Phys.\ Rev.\ D {\bf 70}, 094030 (2004);
%%CITATION = HEP-PH 0408268;%%
R.~Casalbuoni, R.~Gatto, M.~Mannarelli, G.~Nardulli and M.~Ruggieri,
%``Meissner masses in the gCFL phase of QCD,''
Phys.\ Lett.\ B {\bf 605}, 362 (2005);
%%CITATION = HEP-PH 0410401;%%
I.~Giannakis and H.-C.~Ren,
%``Chromomagnetic instability and the LOFF state in a two flavor color
%superconductor,''
Phys.\ Lett.\ B {\bf 611}, 137 (2005);
%%CITATION = HEP-PH 0412015;%%
M.~Alford and Q.~H.~Wang,
%``Photons in gapless color-flavor-locked quark matter,''
J.\ Phys.\ G {\bf 31}, 719 (2005).
%%CITATION = HEP-PH 0501078;%%

\bibitem{BFGOS} D.~Blaschke, S.~Fredriksson, H.~Grigorian, 
A.M.~\"{O}zta\c{s}, and F.~Sandin, 
%``The phase diagram of three-flavor quark matter under compact star
%constraints,''
hep-ph/0503194.
%%CITATION = HEP-PH 0503194;%%


\end{thebibliography}
\end{document}